\documentclass[%
 reprint,
superscriptaddress,
nofootinbib,
 amsmath,amssymb,
 prd,
showkeys]{revtex4-2}

\usepackage[margin=5pt,caption=false]{subfig}
\usepackage{slantsc}
\usepackage{comment}
\usepackage{longtable} 
\usepackage[online]{threeparttablex}	
\usepackage{threeparttable}
\usepackage{booktabs}
\usepackage[figuresright]{rotating}
\usepackage{textcase}
\usepackage{multirow}
\newcommand{\be}{\begin{equation}}
\newcommand{\ee}{\end{equation}}
\newcommand{\bea}{\begin{eqnarray}}
\newcommand{\eea}{\end{eqnarray}}
\newcommand{\hunit}{$\rm{km \ s^{-1} \ Mpc^{-1}}$}
\newcommand{\lcdm}{$\Lambda$CDM}
\newcommand{\pcdm}{$\phi$CDM}

\usepackage{array}

\newcommand{\hii}{H\,\textsc{ii}}
\newcommand{\Om}{\Omega_{m0}}
\newcommand{\Ok}{\Omega_{k0}}
\newcommand{\om}{$\Omega_{m0}$}
\newcommand{\ok}{$\Omega_{k0}$}
\newcommand{\wx}{$w_{\rm X}$}
\newcommand{\wX}{w_{\rm X}}

\newcommand{\mii}{Mg\,\textsc{ii}}

\newcommand{\civ}{C\,\textsc{iv}}

\newcommand{\Feii}{Fe\,\textsc{ii}}
\newcommand{\obh}{\Omega_{b}h^2}
\newcommand{\och}{\Omega_{c}h^2}
\newcommand{\onh}{\Omega_{\nu}h^2}
\newcommand{\obhs}{$\Omega_{b}h^2$}
\newcommand{\ochs}{$\Omega_{c}h^2$}

\usepackage[T1]{fontenc}
\usepackage{scalerel}
\usepackage{tikz}
\usetikzlibrary{svg.path}

\definecolor{orcidlogocol}{HTML}{A6CE39}
\tikzset{
  orcidlogo/.pic={
    \fill[orcidlogocol] svg{M256,128c0,70.7-57.3,128-128,128C57.3,256,0,198.7,0,128C0,57.3,57.3,0,128,0C198.7,0,256,57.3,256,128z};
    \fill[white] svg{M86.3,186.2H70.9V79.1h15.4v48.4V186.2z}
                 svg{M108.9,79.1h41.6c39.6,0,57,28.3,57,53.6c0,27.5-21.5,53.6-56.8,53.6h-41.8V79.1z M124.3,172.4h24.5c34.9,0,42.9-26.5,42.9-39.7c0-21.5-13.7-39.7-43.7-39.7h-23.7V172.4z}
                 svg{M88.7,56.8c0,5.5-4.5,10.1-10.1,10.1c-5.6,0-10.1-4.6-10.1-10.1c0-5.6,4.5-10.1,10.1-10.1C84.2,46.7,88.7,51.3,88.7,56.8z};
  }
}

\newcommand\orcidicon[1]{\href{https://orcid.org/#1}{\mbox{\scalerel*{
\begin{tikzpicture}[yscale=-1,transform shape]
\pic{orcidlogo};
\end{tikzpicture}
}{|}}}}

\usepackage{graphicx}	% Including figure files
\usepackage{amsmath,bm}	% Advanced maths commands
\usepackage{amssymb}	% Extra maths symbols
\usepackage{fnpct}
\usepackage[colorlinks=true,allcolors=blue]{hyperref}
\begin{document}

\preprint{APS/123-QED}

\title{Standardizing a larger, higher-quality, homogeneous sample of reverberation-mapped H$\beta$ active galactic nuclei using the broad-line region radius--luminosity relation}

\author{Shulei Cao$^{\orcidicon{0000-0003-2421-7071}}$}
\email{shuleic@mail.smu.edu}
\affiliation{Department of Physics, Kansas State University, 116 Cardwell Hall, Manhattan, KS 66506, USA}
\affiliation{Department of Physics, Southern Methodist University, Dallas, TX 75205, USA}%
\author{Amit Kumar Mandal$^{\orcidicon{0000-0001-9957-6349}}$}
\affiliation{Center for Theoretical Physics, Polish Academy of Sciences, Al. Lotnik\'{o}w 32/46, 02-668 Warsaw, Poland}%
\author{Michal Zaja\v{c}ek$^{\orcidicon{0000-0001-6450-1187}}$}
\affiliation{Department of Theoretical Physics and Astrophysics, Faculty of Science, Masaryk University, Kotlá\v{r}ská 2, 611 37 Brno, Czech Republic}%
%\author{Bo\.zena Czerny$^{\orcidicon{0000-0001-5848-4333}}$}%
%\email{bcz@cft.edu.pl}
%\affiliation{Center for Theoretical Physics, Polish Academy of Sciences, Al. Lotnik\'{o}w 32/46, 02-668 Warsaw, Poland}%
\author{Bharat Ratra$^{\orcidicon{0000-0002-7307-0726}}$}%
\email{ratra@phys.ksu.edu}
\affiliation{Department of Physics, Kansas State University, 116 Cardwell Hall, Manhattan, KS 66506, USA}%
% \affiliation{ 
% Department of Physics, Kansas State University, 116 Cardwell Hall, Manhattan, KS 66506, USA%\\This line break forced with \textbackslash\textbackslash
% }%

\date{\today}% It is always \today, today,

\begin{abstract}
We present a high-quality, homogeneous sample of 157 H$\beta$ reverberation-mapped active galactic nuclei (RM AGNs) spanning redshifts $0.00308 \leq z \leq 0.8429$, which is approximately 3.8 times larger than the previously available high-quality homogeneous sample. Using the broad-line region radius$-$luminosity relation ($R-L$), which involves the broad H$\beta$ line time delay and the monochromatic luminosity at 5100\,\AA\,, we show that the sample is standardizable by using six spatially flat and nonflat cosmological models. The inferred cosmological model parameters are consistent within 2$\sigma$ uncertainties with those from better established baryon acoustic oscillation and Hubble parameter measurements, with the exception of two nonflat models that are ruled out by other data. The $R-L$ relation slope is found to be flatter ($\gamma=0.428 \pm 0.025$ in the flat $\Lambda$CDM model) than the slope of 0.5 expected from a simple photoionization model as well as the slope found previously for the smaller homogeneous sample. In addition, we find a mild dependence of H$\beta$ $R-L$ relation parameters as well as its intrinsic scatter on the Eddington ratio by comparing the $R-L$ relations for low- and high-accreting equal-sized subsamples. A future analysis of a larger homogeneous sample containing a broader range of luminosities and Eddington ratios is necessary to confirm the standardizability of H$\beta$ AGNs. 

\end{abstract}

 %\keywords{quasars: emission lines; reverberation mapping; cosmological parameters; cosmology: observations; dark energy}

\maketitle

%\tableofcontents
\section{Introduction}

The current standard model of cosmology, the spatially flat $\Lambda$CDM model \citep{peeb84}, where dark energy is represented by the cosmological constant $\Lambda$ and non-relativistic matter is dominated by cold dark matter (CDM), can successfully accommodate the majority of observed characteristics of the Universe at both low and high redshifts \citep[e.g.][]{Farooq_Ranjeet_Crandall_Ratra_2017,scolnic_et_al_2018,planck2018b,eBOSSL_2021}. However, differences in the values of a few $\Lambda$CDM model cosmological parameters inferred from different observational probes have attracted attention, especially differences in the Hubble constant \citep{PerivolaropoulosSkara2021, Abdallaetal2022, DiValentino2025}. At the same time current data cannot rule out mild dark energy dynamics, leaving space for cosmological models beyond the standard model \citep[e.g.][]{CaoRatra2023, deCruzPerez:2024shj, DESI_2025_2025JCAP...02..021A, ParkdeCruzPerezRatra2024}.

It is unclear whether these differences are a consequence of underestimated systematic uncertainties in one or more of the observational probes, e.g., for the Hubble constant ($H_{0}$), in some of the lower-redshift cosmic-ladder measurements \citep[e.g.][]{Chenetal2024, Freedmanetal2025} or in the higher-redshift cosmic microwave background (CMB) measurements, or whether spatially flat $\Lambda$CDM is in fact not the correct cosmological model. Since the difference in Hubble constant values arises when comparing constraints from some nearby cosmic-ladder measurements with those from higher-redshift probes (CMB, baryon acoustic oscillations -- BAO), an insight into these differences can potentially be achieved by employing alternate cosmological probes, especially those at intermediate redshifts that bridge nearby and more distant probes. 

To this end, the standardizability of a variety of astronomical objects has been studied. These include
\begin{itemize}
    \item luminosity versus velocity dispersion correlations of \hii\ starburst galaxies that reach to redshift $z = 2.5$ \citep{Mania_2012,GonzalezMoran2019,Gonzalez-Moran:2021drc,CaoRyanRatra2020, CaoRyanRatra2022,Johnsonetal2022,Mehrabietal2022},
    \item quasar (QSO) angular size measurements that reach to $z = 2.7$ \citep{2017A&A...606A..15C,2019MNRAS.488.3844R,CaoRyanRatra2020,CaoRyanRatra2022,CaoRyanRatra2021,Lianetal2021,Zhengetal2021},
    \item QSO X-ray and UV flux measurements of sources that reach to $z = 7.5$ \citep{RisalitiLusso2019,KhadkaRatra2020a,KhadkaRatra2020b,KhadkaRatra2021,Lussoetal2020,Yang_Banerjee_Colgain_2020,Rezaeietal2021,ZhaoXia2021,KhadkaRatra2022},
    \item $\gamma$-ray bursts (GRBs) that reach to $z = 8.2$ \citep{Wang_2016,Dirirsa2019,Huetal_2021,Luongoetal2021,CaoRatra2022,Favaleetal2024,Lietal2024,Bargiacchietal2025,Dengetal2025,CaoRatra2025},
    \item QSO parallax distance measurements by combining spetroastrometry and reverberation mapping (RM) \citep{2018Natur.563..657G,2020NatAs...4..517W},
    \item super-Eddington QSOs \citep{2013PhRvL.110h1301W,2014MNRAS.442.1211M,2019Atoms...7...18M},
    \item gravitational-wave sources \citep{2017ApJ...851L..36G,2017Natur.551...85A,2024AnP...53600180M,2025arXiv250614150Z}.
\end{itemize}
Therefore, testing the standardizability of a sample requires verifying whether or not the assumed correlations (discussed in detail below) are independent of the adopted cosmological model. In practice this requires simultaneously constraining the parameters of the correlation and the cosmological parameters of a number of different cosmological models. This approach \cite{KhadkaRatra2020c, Caoetal_2021} circumvents the circularity problem inherent in such analyses.

This technique has already been applied to assess the standardizability of GRB data compilations \cite{Khadkaetal_2021b, CaoKhadkaRatra2022, CaoDainottiRatra2022, CaoDainottiRatra2022b, CaoRatra2024b, CaoRatra2025}, assuming the Amati correlation between the rest-frame peak energy and the rest-frame isotropic radiated energy \cite{Amatietal2002} as well as a number of other correlations. It shows that a more homogeneous current compilation of Amati-correlated GRBs can be standardized and is the largest available standardizable sample of GRBs \cite{Khadkaetal_2021b}.   

It has also been used to show that the \hii\ starburst galaxy luminosity versus velocity dispersion correlation is redshift-dependent and thus non-standardizable for the current \hii\ galaxy dataset of \citet{Gonzalez-Moran:2021drc}, rendering it unsuitable for cosmological applications \cite{CaoRatra2024a, Melnick:2024ywi}.

Several methods listed above employ QSOs that are attractive alternate probes since they cover a large redshift range from the nearby Universe to $z\gtrsim 6$. However, the above technique \cite{KhadkaRatra2020c, Caoetal_2021} has also been used to demonstrate that the \citet{Lussoetal2020} $L_X-L_{UV}$ correlation QSO dataset is not suitable for cosmology due to its lack of standardizability \cite{KhadkaRatra2021, KhadkaRatra2022, Petrosian:2022tlp, Khadka:2022aeg, Lietal2025, Wuetal2025}. This lack of standardizability was suggested to be caused by (potentially host-galaxy-dependent) dust extinction that affects in different ways the X-ray and UV flux measurements, which makes it challenging to mitigate the extinction effect \citep{Zajaceketal2024}.  

Another way to standardize QSOs is to use the broad-line region (BLR) radius$-$luminosity ($R-L$) relation \citep{2013peag.book.....N,MartinezAldama2019,2021bhns.confE...1K,Czerny:2022xfj}. In the original form, this is the power-law relation between the monochromatic luminosity of the QSO and the radial distance of the BLR from the inner accretion disk surrounding the supermassive black hole (SMBH), corresponding to the region where broad emission lines originate. Instead of the linear distance or radius, it is more convenient to express the extent as the time delay ($\tau$) and the two quantities are related simply via the speed of light, i.e., $R=c\tau$. After the observational establishment of the active galactic nucleus (AGN)\footnote{We use AGN here instead of QSO since the H$\beta$ sources are typically of lower absolute luminosity than the \mii\ and \civ\ QSO sources.} H$\beta$ $R-L$ relation \citep{2013ApJ...767..149B}, an $R-L$ relation was also confirmed to exist for broad UV lines in the rest-frame QSO spectrum, specifically for \mii\ \citep{Zajaceketal2021,Yuetal2021,2024SSRv..220...29Z} and \civ\ lines \citep{2021ApJ...915..129K}. While the H$\beta$ $R-L$ relation observationally predominantly holds for lower-redshift QSOs, \mii\ and \civ\ $R-L$ relations are observationally established for higher-redshift QSOs since in their case these lines are redshifted to the observer's optical domain. In the early works studying the correlation between the monochromatic luminosity and the rest-frame BLR time-delay, the $R-L$ relation contained just two parameters ($\tau$, $L$) and exhibited a relatively small intrinsic scatter at the level of $\sim 0.1-0.2$ dex \citep{2013ApJ...767..149B}. When the number of sources increased thanks to intensive surveys such as the Sloan Digital Sky Survey Reverberation Mapping program (SDSS-RM), which also included the sources with substantially higher Eddington ratios, the scatter in the $R-L$ relation increased considerably \citep[see e.g.][]{2016ApJ...825..126D}, with the general trend of time-delay shortening for higher-accreting sources \citep{MartinezAldama2019}. This has shown that the $R-L$ relation could have an extra dependence on the parameters that are related to the relative accretion rate of the sources, which would make $R-L$ relations more complex (multidimensional), thus deviating from simple two-parameter power laws \citep[see e.g.][and references therein]{2020ApJ...903...86M}. However, including the Eddington ratio and related quantities as the third parameter is problematic since it depends on the monochromatic luminosity and the virialized SMBH mass, which enhances the correlation artificially. It has been suggested to use other accretion-rate proxies that are independent, such as the fractional variability or the relative \Feii\ strength \citep{2019ApJ...886...42D,2020ApJ...903...86M,Zajaceketal2021}. However, it is also not yet clear whether the scatter is caused primarily by the relative accretion rate or another parameter \citep{2017ApJ...851...21G}, such as the UV/optical spectral energy distribution shape and the relative contribution of ionizing radiation \citep{2020ApJ...899...73F}. Moreover, the scatter could be affected by the AGN sample selection due to a specific monitoring cadence \citep{2017ApJ...851...21G}. In addition, it was found that including the third parameter in the $R-L$ relation does not significantly impact the $R-L$ intrinsic scatter and the cosmological-parameter constraints when they are allowed to vary \citep{Khadkaetal2021c,Khadkaetal2022a}. We stress that most of the previous results concerning the $R-L$ relation were obtained for the fixed flat $\Lambda$CDM model. 

To verify whether H$\beta$ AGNs, and \mii\ and \civ\ QSOs are standardizable, it is necessary to check whether the normalization and the slope parameters of the $R-L$ relation are independent of the adopted cosmological model. It is also necessary to avoid the circularity problem, i.e. the $R-L$ relation parameters and the parameters of the adopted cosmological model need to be constrained simultaneously. This requires studying the $R-L$ relation in multiple cosmological models. For a detailed description of the Markov chain Monte Carlo (MCMC) parameter inference, see, e.g., \citep{Khadkaetal_2021a} and \citep{Cao:2022pdv}. \citet{Khadkaetal_2021a} found that a sample of 78 \mii\ QSOs in the redshift range $0.0033 \lesssim z \lesssim 1.89$ is standardizable and the resulting cosmological model parameter constraints, albeit weak, are consistent with the standard spatially flat $\Lambda$CDM model, but do not exclude models with a little spatial curvature or mild dark energy dynamics. Similarly, 38 \civ\ QSOs in the redshift range $0.001 \lesssim z \lesssim 3.37$ were found to be standardizable and the inferred cosmological constraints were found to be consistent with those coming from standard cosmological probes, specifically BAO and Hubble parameter [$H(z)$] data \citep{Cao:2022pdv}. When these \civ\ data are used jointly with \mii\ and $H(z)$ + BAO data, the cosmological parameter constraints become slightly tighter, by $\sim 0.1 \sigma$, than for $H(z)$ + BAO data alone.

In contrast, although the lower-redshift H$\beta$ AGNs (118 AGNs in the redshift range $0.0023 \lesssim z \lesssim 0.89$) were also found to be standardizable, the cosmological parameter constraints favored currently decelerated cosmological expansion and were in $\sim 2\sigma$ tension with the constraints inferred using standard cosmological probes \citep{Khadkaetal2021c}. The tension was attributed to the  heterogeneity of the sample, as the time delays were determined using different methodologies, potentially introducing systematic biases across the subsamples.\footnote{See Ref.\ \citep{Caoetal2024} for a similar result from a heterogeneous sample of \mii\ and \civ\ QSOs.}

This motivated the compilation of a homogeneous, good-quality H$\beta$ AGN sample to see whether the tension persists. In \citet{Caoetal2025}, 41 AGNs in the redshift range $0.00415 \lesssim z \lesssim 0.474$ were found to be standardizable using four $R-L$ relations involving H$\alpha$ and H$\beta$ broad lines and monochromatic and broad H$\alpha$ luminosities. Cosmological parameter constraints were found to be weak and within $\lesssim 2\sigma$, consistent with those inferred using standard cosmological probes. Hence, the homogeneity of the sample helped to decrease the tension ($\lesssim 2\sigma$ in comparison with $\gtrsim 2\sigma$ for the 118 AGN sample).  

In this paper, we use a sample of 157 higher-quality H$\beta$ AGNs in the redshift range $0.00308 \lesssim z \lesssim 0.8429$, which were compiled by \citep{2024ApJS..275...13W} using several time delay quality criteria. This sample was reanalyzed using a homogeneous methodology and it is significantly larger by a factor of $\sim 3.8$ with respect to the previous small homogeneous sample of 41 sources. We find that this sample is standardizable using the H$\beta$ $R-L$ relation and this $R-L$ relation slope is flatter than that the slope of $0.5$ expected from simple photoionization arguments\footnote{ The slope is derived simply from the relation for the ionization parameter $U$ for BLR clouds located at distance $R$ from the central photoionizing source, under the assumption of equilibrium between the photoionization rate and the recombination rate: $U=Q_{\rm ion}(H)/(4\pi R^2 c n_{\rm e})$, where $Q_{\rm ion}(H)=\int_{\nu_{\rm i}}^{+\infty} L_{\nu}/(h\nu)\mathrm{d}\nu$ is the rate of hydrogen-ionizing photons integrated over the monochromatic luminosity of the central source $L_{\nu}$ and $n_{\rm e}$ is the electron density of BLR clouds that is proportional to the recombination rate \citep[see e.g.][for more discussion]{2021bhns.confE...1K}. Assuming that $4\pi c Un_{\rm e}\approx \text{const}$ for BLR clouds across AGNs, we get $R^2=KQ_{\rm ion}(H)$, from which $R\propto L_{\nu}^{1/2}$. We stress that the assumption of a constant $Un_{\rm e}$ may not hold universally and hence the photoionization derivation of the slope of 0.5 is only approximate. It can, however, still serve as a useful value for comparison with the observationally inferred slope of the $R-L$ relation since the early results for lower-luminosity AGNs were in agreement with this value within uncertainties \citep[see e.g.][]{2013ApJ...767..149B}.} and also flatter than that of the previous 41 source sample. This can be attributed to the 157 source sample containing a larger fraction of higher-accreting sources, for which the time delay is smaller for a given monochromatic luminosity. As for the marginalized 1D cosmological parameter constraints, there are within $\lesssim 2\sigma$ consistent with those found for standard cosmological probes. However, we find $\gtrsim 2\sigma$ differences in marginalized 2D cosmological parameter contours for the non-flat $\Lambda$CDM model and the non-flat XCDM parametrization. This indicates a tension that complicates joint analyses of RM H$\beta$ QSOs and $H(z)$ + BAO data in these two cases.

This manuscript is structured as follows. In Sec.~\ref{sec:model}, we introduce the six cosmological models used in this study. We list the datasets used for constraining $R-L$ relation and cosmological model parameters in Sec.~\ref{sec:data}. Subsequently, in Sec.~\ref{sec:analysis} we provide details of the parameter inference methodology. We describe the main results, specifically concerning cosmological model and $R-L$ relation parameters, in Sec.~\ref{sec:results}. Finally, we summarize the main findings in Sec.~\ref{sec:conclusion}.

\section{Cosmological models}
\label{sec:model}

To determine whether H$\beta$ RM AGN sources can be standardized using the $R-L$ relation, we simultaneously constrain the parameters of this relation and the cosmological parameters of six spatially flat and nonflat relativistic dark energy cosmological models. The premise is that if the derived $R-L$ relation parameters remain consistent across different assumed cosmological models, then the AGN dataset can be considered to be standardizable. This approach \cite{KhadkaRatra2020c, Caoetal_2021} circumvents the circularity problem inherent in such analyses.

In each cosmological model, the luminosity distance $D_L(z)$ is predicted as a function of redshift $z$ and the cosmological model parameters through the dimensionless Hubble parameter $E(z) = H(z)/H_0$,
\begin{equation}
  \label{eq:DL}
\resizebox{0.475\textwidth}{!}{%
    $D_L(z) = 
    \begin{cases}
    \frac{c(1+z)}{H_0\sqrt{\Omega_{\rm k0}}}\sinh\left[\frac{H_0\sqrt{\Omega_{\rm k0}}}{c}D_C(z)\right] & \text{if}\ \Omega_{\rm k0} > 0, \\
    \vspace{1mm}
    (1+z)D_C(z) & \text{if}\ \Omega_{\rm k0} = 0,\\
    \vspace{1mm}
    \frac{c(1+z)}{H_0\sqrt{|\Omega_{\rm k0}|}}\sin\left[\frac{H_0\sqrt{|\Omega_{\rm k0}|}}{c}D_C(z)\right] & \text{if}\ \Omega_{\rm k0} < 0,
    \end{cases}$%
    }
\end{equation}
where the comoving distance is given by
\begin{equation}
\label{eq:gz}
   D_C(z) = \frac{c}{H_0}\int^z_0 \frac{dz'}{E(z')}.
\end{equation}
Here $c$ represents the speed of light, $H_0$ denotes the Hubble constant, and \ok\ is the current spatial curvature density parameter.\footnote{For recent discussions on the implications of and constraints on spatial curvature, refer to Refs.\ \cite{Oobaetal2018b, ParkRatra2019b, DiValentinoetal2021a, ArjonaNesseris2021, Dhawanetal2021, Renzietal2021, Gengetal2022, MukherjeeBanerjee2022, Glanvilleetal2022, Wuetal2023, deCruzPerezetal2023, DahiyaJain2022, Stevensetal2023,Favaleetal2023, Qietal2023, Shimon:2024mbm, Wu:2024faw}.} We adopt a standard model for the neutrino sector, assuming one massive and two massless species, which sets the effective number of relativistic neutrino species to $N_{\rm eff} = 3.046$ and the total neutrino mass to $\sum m_{\nu} = 0.06$ eV. This implies a present energy density parameter for nonrelativistic neutrinos, $\onh=\sum m_{\nu}/(93.14\ \rm eV)$, where $h$ is the Hubble constant in units of 100 \hunit. The total present nonrelativistic matter density parameter, $\Om = (\onh + \obh + \och)/{h^2}$, encompasses contributions from neutrinos, baryonic matter, and cold dark matter, while the influence of radiation at late times is negligible and thus ignored in our study.

We explore six different cosmological models. In the standard \lcdm\ framework, dark energy is attributed to a cosmological constant, $\Lambda$, corresponding to a fluid with equation of state parameter $w_{\rm DE}=-1$. An alternative to the cosmological constant is the XCDM parametrization of dark energy, which models dynamical dark energy as a spatially homogeneous fluid with a constant equation of state parameter $w_{\rm DE}$ that can deviate from $-1$. It is important to acknowledge that XCDM is a phenomenological description and does not provide a complete physical model for dark energy inhomogeneities. For both \lcdm\ and XCDM, the evolution of the cosmological expansion rate is encapsulated in the dimensionless Hubble parameter as
\be
\label{eq:EzL}
\resizebox{0.475\textwidth}{!}{%
    $E(z) = \sqrt{\Om\left(1 + z\right)^3 + \Ok\left(1 + z\right)^2 + \Omega_{\rm DE0}\left(1+z\right)^{3(1+w_{\rm DE})}},$%
    }
\ee
where $\Omega_{\rm DE0} = 1 - \Om - \Ok$ represents the present dark energy density parameter. Due to the weak constraints on $H_0$ and $\Omega_{b}$ from RM AGN data alone, we fix these parameters to $H_0=70$ \hunit\ and $\Omega_{b}=0.05$ for the AGN only analyses. Consequently, in this case, our free parameters are $\{\Om,\Ok\}$ in \lcdm\ and $\{\Om,\Ok,\wX\}$ in XCDM (with $\Ok=0$ for the flat cases). For analyses incorporating $H(z)$ + BAO data, we adopt $\{H_0, \obh, \och\}$ in place of $\{\Om\}$ as fundamental parameters.

We also consider the \pcdm\ models \citep{peebrat88,ratpeeb88,pavlov13}\footnote{For recent studies constraining \pcdm, see Refs.\ \cite{ooba_etal_2018b, ooba_etal_2019, park_ratra_2018, park_ratra_2019b, park_ratra_2020, Singhetal2019, UrenaLopezRoy2020, SinhaBanerjee2021, deCruzetal2021, Xuetal2022, Jesusetal2022, Adiletal2023, Dongetal2023, VanRaamsdonkWaddell2023, Avsajanishvilietal2024, VanRaamsdonkWaddell2024a, Thompson2024}.}, where dynamical dark energy arises from a time- and space-dependent scalar field $\phi$ governed by an inverse power law potential energy density
\be
\label{PE}
V(\phi)=\frac{1}{2}\kappa m_p^2\phi^{-\alpha}.
\ee
The evolution of the Universe in these models is described by the Friedmann equation
\be
\label{eq:Ezp}
    E(z) = \sqrt{\Om\left(1 + z\right)^3 + \Ok\left(1 + z\right)^2 + \Omega_{\phi}(z,\alpha)},
\ee
where
\be
\label{Op}
\Omega_{\phi}(z,\alpha)=\frac{1}{6H_0^2}\bigg[\frac{1}{2}\dot{\phi}^2+V(\phi)\bigg],
\ee
is the dynamical scalar field dark energy density parameter, determined numerically through the Friedmann equation and the scalar field's equation of motion
\be
\label{em}
\ddot{\phi}+3H\dot{\phi}+V'(\phi)=0.
\ee
In these equations, $\alpha$ is the exponent of the inverse power law potential energy density with $\alpha = 0$ corresponding to \lcdm, $m_p$ is the Planck mass, and time and scalar field derivatives are denoted by overdots and primes, respectively. The normalization constant $\kappa$ is determined using the shooting method within the Cosmic Linear Anisotropy Solving System (\textsc{class}) code \citep{class}. For RM AGN data analyses in \pcdm, $\{\Om,\Ok,\alpha\}$ are treated as free parameters (with $\Ok=0$ in the flat case). When $H(z)$ + BAO data are included, $\Om$ is treated as a derived parameter, with $\{H_0, \obh, \och\}$ used as inputs instead.

\begin{figure}
\centering
\includegraphics[width=\columnwidth]{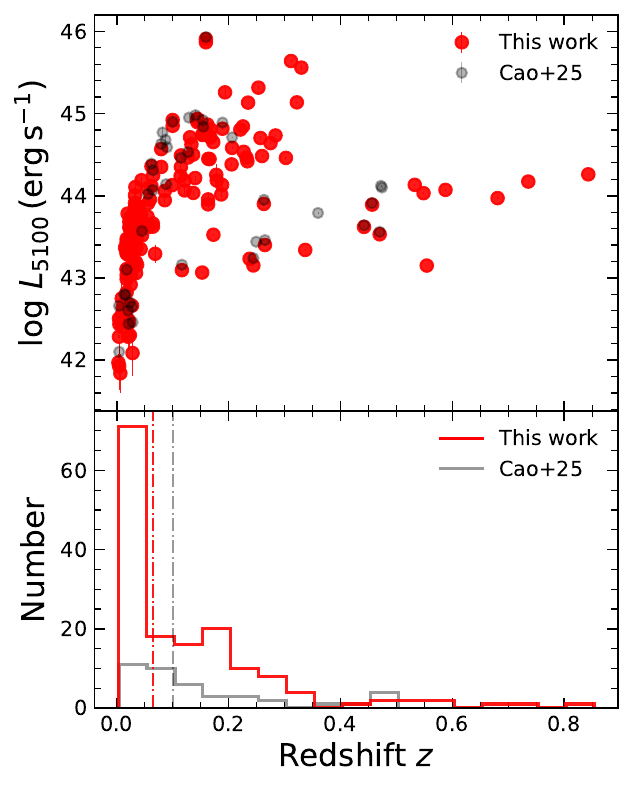}
\caption{The characteristics of the selected Best sample of H$\beta$-reverberation mapped AGNs (shown in red) compared with those of the previous H$\alpha$ and H$\beta$ reverberation-mapped sample (shown in black) from \citet{Caoetal2025}. The top panel displays the continuum luminosity at 5100 {\AA} as a function of redshift, while the bottom panel shows the redshift distributions in histogram form, using the same uniform bin size for both samples. The dashed vertical line indicates the median redshift of the sample. The continuum luminosities at 5100 {\AA} are derived from the monochromatic flux at 5100 {\AA}, assuming a flat $\Lambda$CDM cosmological model with $H_0=72\,{\rm km\,s^{-1}\,Mpc^{-1}}$ and $\Omega_{ m0}=0.3$. Compared to the previous sample, the current sample spans a broader range in both redshift and luminosity.}
\label{fig:dist}
\end{figure}

\section{Sample and Data}
\label{sec:data}

In this section, we primarily focus on the AGN sample selection, which includes a comprehensive set of H$\beta$ reverberation mapping measurements spanning a broad range of redshifts and luminosities. Additionally, we incorporate $H(z)$ and BAO data \citep{CaoRatra2023} to complement our analysis, providing a broader context for the cosmological implications of the $R-L$ relation.

\subsection{AGN Sample and Data}

In this study, we utilize the largest currently available, uniform, compilation of H$\beta$ RM AGN measurements, based on a uniform lag analysis conducted by \citet{2024ApJS..275...13W}. These authors compiled approximately 280 H$\beta$-RM measurements, including H$\beta$ and continuum light curves, from the literature available before 2022 and 32 AGNs from the Seoul National University AGN Monitoring Project \citep[SAMP;][]{2024ApJ...962...67W} to constrain the $R_{\mathrm{H\beta}} - L_{5100}$ relation, where $R_{\mathrm{H\beta}}$ and $L_{5100}$ refer to H$\beta$ BLR radius and optical monochromatic luminosity at 5100 {\AA}, respectively.

A critical step in their analysis is accounting for host-galaxy contamination, which can significantly bias AGN luminosity estimates. Therefore, to accurately investigate and use the $R - L_{5100}$ relation, it is essential to correct for this contamination. Initially, the dataset contained 312 measurements (280 from the literature and 32 from SAMP). However, two AGNs lacked available light curves, 38 measurements did not have host-contamination-corrected AGN luminosities, and 28 measurements were identified as having unreliable H$\beta$ lags. After excluding these cases, the parent sample was reduced to 244 measurements with available host-corrected $L_{5100}$.

\begin{turnpage}
\begin{table*}
\centering
\resizebox{2.7\columnwidth}{!}{%
\begin{threeparttable}
\caption{Properties of the Best sample}
\label{tab:sam_best}
\setlength{\tabcolsep}{18pt}
\begin{tabular}{lcccccr}
\toprule\toprule
Object & $z$ & log F$_{5100}$ ($\mathrm{erg \, s^{-1} \, cm^{-2}}$) & log L$_{5100}$ ($\mathrm{erg \, s^{-1}}$) & $\tau_{\mathrm{H\beta, ICCF}}$ (days) & $\tau_{\mathrm{H\beta, PyROA}}$ (days) & Reference \\
(1) & (2) & (3) & (4) & (5) & (6) & (7) \\
\midrule

Mrk50 & 0.02502 & $ -11.212 \pm 0.09 $ & $ 42.918 \pm 0.09 $ & $ 7.8 _ {-0.9} ^ {+1.3} $ & $ 6.6 _ {-0.4} ^ {+0.4} $ & \citep{2011ApJ...743L...4B, Barth_2015} \\
Mrk40 & 0.02128 & $ -11.539 \pm 0.08 $ & $ 42.447 \pm 0.08 $ & $ 5.2 _ {-0.8} ^ {+0.5} $ & $ 4.9 _ {-0.2} ^ {+0.2} $ & \citep{Barth_2015} \\
PG1310-108 & 0.03534 & $ -11.07 \pm 0.06 $ & $ 43.367 \pm 0.06 $ & $ 6.6 _ {-2.9} ^ {+2.2} $ & $ 10.5 _ {-1.0} ^ {+0.9} $ & \citep{Barth_2015} \\
Zw229-015 & 0.02735 & $ -11.546 \pm 0.06 $ & $ 42.662 \pm 0.06 $ & $ 4.1 _ {-1.0} ^ {+1.4} $ & $ 1.7 _ {-0.4} ^ {+0.4} $ & \citep{Barth_2015} \\
Mrk1511 & 0.03458 & $ -11.36 \pm 0.06 $ & $ 43.058 \pm 0.06 $ & $ 6.4 _ {-0.9} ^ {+1.0} $ & $ 6.8 _ {-0.4} ^ {+0.4} $ & \citep{Barth_2015, 2013ApJ...769..128B} \\
NGC4593 & 0.00946 & $ -10.871 \pm 0.07 $ & $ 42.404 \pm 0.07 $ & $ 3.2 _ {-2.3} ^ {+1.5} $ & $ 2.5 _ {-0.1} ^ {+0.1} $ & \citep{Barth_2015, 2013ApJ...769..128B} \\
NGC4151 & 0.00415 & $ -10.271 \pm 0.02 $ & $ 42.284 \pm 0.02 $ & $ 6.6 _ {-0.5} ^ {+0.5} $ & $ 5.9 _ {-0.4} ^ {+0.4} $ & \citep{2006ApJ...651..775B} \\
Mrk1310 & 0.02077 & $ -11.683 \pm 0.17 $ & $ 42.283 \pm 0.17 $ & $ 3.7 _ {-0.5} ^ {+0.5} $ & $ 3.8 _ {-0.2} ^ {+0.2} $ & \citep{Bentzetal2009} \\
Mrk202 & 0.02359 & $ -11.776 \pm 0.18 $ & $ 42.302 \pm 0.18 $ & $ 5.3 _ {-3.3} ^ {+7.4} $ & $ 2.3 _ {-0.8} ^ {+0.8} $ & \citep{Bentzetal2009} \\
NGC4748 & 0.01575 & $ -11.167 \pm 0.13 $ & $ 42.555 \pm 0.13 $ & $ 5.7 _ {-1.6} ^ {+1.8} $ & $ 4.8 _ {-1.5} ^ {+1.6} $ & \citep{Bentzetal2009} \\
NGC5548 & 0.01788 & $ -10.848 \pm 0.11 $ & $ 42.985 \pm 0.11 $ & $ 4.2 _ {-1.4} ^ {+1.0} $ & $ 4.6 _ {-0.3} ^ {+0.3} $ & \citep{Bentzetal2009} \\
NGC6814 & 0.00451 & $ -10.704 \pm 0.29 $ & $ 41.924 \pm 0.29 $ & $ 6.5 _ {-1.0} ^ {+0.9} $ & $ 6.4 _ {-0.5} ^ {+0.5} $ & \citep{Bentzetal2009} \\
SBS1116+583A & 0.02836 & $ -12.156 \pm 0.28 $ & $ 42.085 \pm 0.28 $ & $ 2.2 _ {-0.5} ^ {+0.9} $ & $ 2.5 _ {-0.4} ^ {+0.5} $ & \citep{Bentzetal2009} \\
Mrk40 & 0.02128 & $ -11.499 \pm 0.11 $ & $ 42.487 \pm 0.11 $ & $ 4.3 _ {-0.5} ^ {+0.5} $ & $ 3.6 _ {-0.2} ^ {+0.2} $ & \citep{Bentzetal2009, 2008ApJ...689L..21B} \\
UGC06728 & 0.00659 & $ -11.12 \pm 0.24 $ & $ 41.839 \pm 0.24 $ & $ 1.1 _ {-0.9} ^ {+0.5} $ & $ 2.1 _ {-0.4} ^ {+0.3} $ & \citep{2016ApJ...831....2B} \\
Mrk290 & 0.03037 & $ -11.168 \pm 0.06 $ & $ 43.133 \pm 0.06 $ & $ 9.6 _ {-1.0} ^ {+1.4} $ & $ 7.7 _ {-0.4} ^ {+0.4} $ & \citep{2010ApJ...721..715D} \\
Mrk817 & 0.0316 & $ -10.553 \pm 0.05 $ & $ 43.784 \pm 0.05 $ & $ 10.1 _ {-2.4} ^ {+2.0} $ & $ 9.4 _ {-1.6} ^ {+1.5} $ & \citep{2010ApJ...721..715D} \\
NGC3227 & 0.00483 & $ -10.253 \pm 0.11 $ & $ 42.435 \pm 0.11 $ & $ 4.3 _ {-1.0} ^ {+1.0} $ & $ 3.1 _ {-0.7} ^ {+0.8} $ & \citep{2010ApJ...721..715D} \\
NGC3516 & 0.00906 & $ -10.486 \pm 0.21 $ & $ 42.751 \pm 0.21 $ & $ 11.4 _ {-1.9} ^ {+1.0} $ & $ 7.1 _ {-0.6} ^ {+0.6} $ & \citep{2010ApJ...721..715D} \\
NGC5548 & 0.01788 & $ -11.138 \pm 0.18 $ & $ 42.695 \pm 0.18 $ & $ 2.3 _ {-0.9} ^ {+0.9} $ & $ 18.2 _ {-0.8} ^ {+0.7} $ & \citep{2010ApJ...721..715D} \\
Mrk704 & 0.02962 & $ -10.748 \pm 0.03 $ & $ 43.532 \pm 0.03 $ & $ 14.3 _ {-1.4} ^ {+1.8} $ & $ 14.9 _ {-0.7} ^ {+0.7} $ & \citep{2018ApJ...866..133D} \\
NGC3516 & 0.00906 & $ -10.686 \pm 0.2 $ & $ 42.551 \pm 0.2 $ & $ 7.6 _ {-1.5} ^ {+1.4} $ & $ 8.0 _ {-0.8} ^ {+0.7} $ & \citep{2018ApJ...866..133D} \\
NGC4151 & 0.00415 & $ -10.051 \pm 0.06 $ & $ 42.504 \pm 0.06 $ & $ 6.8 _ {-0.5} ^ {+0.5} $ & $ 6.8 _ {-0.2} ^ {+0.2} $ & \citep{2018ApJ...866..133D} \\
NGC5548 & 0.01788 & $ -10.728 \pm 0.06 $ & $ 43.105 \pm 0.06 $ & $ 2.7 _ {-0.9} ^ {+0.6} $ & $ 3.6 _ {-0.3} ^ {+0.4} $ & \citep{2018ApJ...866..133D} \\
3C390.3 & 0.0559 & $ -11.231 \pm 0.1 $ & $ 43.617 \pm 0.1 $ & $ 12.4 _ {-7.3} ^ {+6.4} $ & $ 13.0 _ {-3.6} ^ {+10.6} $ & \citep{1998ApJS..115..185D, 2004ApJ...613..682P} \\
Mrk142 & 0.04521 & $ -11.141 \pm 0.05 $ & $ 43.515 \pm 0.05 $ & $ 5.6 _ {-1.4} ^ {+1.5} $ & $ 7.5 _ {-1.0} ^ {+1.1} $ & \citep{2014ApJ...782...45D} \\
IRASF12397+3333 & 0.04439 & $ -10.452 \pm 0.04 $ & $ 44.188 \pm 0.04 $ & $ 11.5 _ {-1.9} ^ {+2.8} $ & $ 21.7 _ {-1.7} ^ {+2.3} $ & \citep{2014ApJ...782...45D, 2015ApJ...804..138H} \\
Mrk335 & 0.02461 & $ -10.527 \pm 0.04 $ & $ 43.589 \pm 0.04 $ & $ 9.8 _ {-1.4} ^ {+1.4} $ & $ 10.2 _ {-0.9} ^ {+1.0} $ & \citep{2014ApJ...782...45D, 2015ApJ...804..138H} \\
J075101.42+291419.1 & 0.12141 & $ -11.496 \pm 0.05 $ & $ 44.064 \pm 0.05 $ & $ 27.8 _ {-3.9} ^ {+4.9} $ & $ 29.9 _ {-2.3} ^ {+3.2} $ & \citep{2015ApJ...806...22D} \\
J080101.41+184840.7 & 0.14027 & $ -11.482 \pm 0.03 $ & $ 44.215 \pm 0.03 $ & $ 9.1 _ {-3.9} ^ {+7.3} $ & $ 5.5 _ {-2.4} ^ {+1.6} $ & \citep{2015ApJ...806...22D} \\
J081441.91+212918.5 & 0.16338 & $ -11.887 \pm 0.07 $ & $ 43.955 \pm 0.07 $ & $ 18.9 _ {-9.1} ^ {+8.5} $ & $ 21.7 _ {-1.2} ^ {+1.3} $ & \citep{2015ApJ...806...22D} \\
J093922.89+370943.9 & 0.18688 & $ -11.955 \pm 0.04 $ & $ 44.015 \pm 0.04 $ & $ 10.7 _ {-3.3} ^ {+3.3} $ & $ 14.1 _ {-1.7} ^ {+1.4} $ & \citep{2015ApJ...806...22D} \\
J102339.64+523349.6 & 0.13695 & $ -11.64 \pm 0.03 $ & $ 44.034 \pm 0.03 $ & $ 26.6 _ {-8.0} ^ {+5.5} $ & $ 24.2 _ {-4.8} ^ {+5.2} $ & \citep{2016ApJ...825..126D} \\
J074352.02+271239.5 & 0.25334 & $ -10.952 \pm 0.02 $ & $ 45.315 \pm 0.02 $ & $ 65.3 _ {-17.8} ^ {+8.1} $ & $ 47.3 _ {-1.8} ^ {+1.8} $ & \citep{2018ApJ...856....6D} \\
J075101.42+291419.1 & 0.12141 & $ -11.376 \pm 0.04 $ & $ 44.184 \pm 0.04 $ & $ 21.1 _ {-5.4} ^ {+6.5} $ & $ 19.2 _ {-2.0} ^ {+1.9} $ & \citep{2018ApJ...856....6D} \\
J075949.54+320023.8 & 0.18836 & $ -11.846 \pm 0.06 $ & $ 44.132 \pm 0.06 $ & $ 18.8 _ {-5.7} ^ {+11.3} $ & $ 12.3 _ {-1.4} ^ {+1.4} $ & \citep{2018ApJ...856....6D} \\
J081441.91+212918.5 & 0.16338 & $ -11.947 \pm 0.04 $ & $ 43.895 \pm 0.04 $ & $ 23.4 _ {-5.2} ^ {+7.9} $ & $ 29.7 _ {-2.9} ^ {+2.9} $ & \citep{2018ApJ...856....6D} \\
J083553.46+055317.1 & 0.2054 & $ -11.68 \pm 0.02 $ & $ 44.381 \pm 0.02 $ & $ 28.6 _ {-4.6} ^ {+5.6} $ & $ 25.5 _ {-3.1} ^ {+3.5} $ & \citep{2018ApJ...856....6D} \\
J084533.28+474934.5 & 0.30251 & $ -11.984 \pm 0.02 $ & $ 44.46 \pm 0.02 $ & $ 21.7 _ {-4.2} ^ {+8.9} $ & $ 18.4 _ {-0.9} ^ {+1.0} $ & \citep{2018ApJ...856....6D} \\
J093302.68+385228.0 & 0.17802 & $ -11.669 \pm 0.13 $ & $ 44.254 \pm 0.13 $ & $ 20.9 _ {-4.9} ^ {+4.5} $ & $ 18.2 _ {-2.5} ^ {+2.4} $ & \citep{2018ApJ...856....6D} \\

\bottomrule\bottomrule
\end{tabular}
%\begin{tablenotes}
%\item Columns are: (1) object name, (2) redshift corrected for peculiar velocity, (3)  flux at 5100 {\AA} in log scale, (4) luminosity at 5100 {\AA} in log scale, (5) rest-frame H$\beta$ time lag from ICCF, (6) rest-frame H$\beta$ time lag from {\tt PyROA}, and (7) references. Note that luminosities are derived for the flat $\Lambda$CDM model ($H_0=72\,{\rm km\,s^{-1}\,Mpc^{-1}}$ and $\Omega_{ m0}=0.3$).
%\end{tablenotes}
\end{threeparttable}%
}
\end{table*}
\end{turnpage}

\begin{turnpage}
\begin{table*}
\centering
\resizebox{2.7\columnwidth}{!}{%
\begin{threeparttable}
%\contcaption{}
\setlength{\tabcolsep}{18pt}
\begin{tabular}{lcccccr}
\toprule\toprule
Object & $z$ & log F$_{5100}$ ($\mathrm{erg \, s^{-1} \, cm^{-2}}$) & log L$_{5100}$ ($\mathrm{erg \, s^{-1}}$) & $\tau_{\mathrm{H\beta, ICCF}}$ (days) & $\tau_{\mathrm{H\beta, PyROA}}$ (days) & Reference \\
(1) & (2) & (3) & (4) & (5) & (6) & (7) \\
\midrule

J101000.68+300321.5 & 0.25702 & $ -11.579 \pm 0.02 $ & $ 44.702 \pm 0.02 $ & $ 38.4 _ {-6.6} ^ {+11.2} $ & $ 22.1 _ {-2.1} ^ {+2.3} $ & \citep{2018ApJ...856....6D} \\
3C382 & 0.05512 & $ -11.089 \pm 0.07 $ & $ 43.746 \pm 0.07 $ & $ 38.5 _ {-6.4} ^ {+4.6} $ & $ 40.3 _ {-1.3} ^ {+1.4} $ & \citep{2017ApJ...840...97F} \\
MCG+08-11-011 & 0.02051 & $ -10.673 \pm 0.04 $ & $ 43.281 \pm 0.04 $ & $ 15.6 _ {-0.5} ^ {+0.8} $ & $ 15.0 _ {-0.3} ^ {+0.3} $ & \citep{2017ApJ...840...97F} \\
Mrk374 & 0.04283 & $ -10.854 \pm 0.04 $ & $ 43.754 \pm 0.04 $ & $ 14.8 _ {-5.8} ^ {+3.4} $ & $ 12.8 _ {-1.6} ^ {+1.5} $ & \citep{2017ApJ...840...97F} \\
NGC2617 & 0.01517 & $ -11.022 \pm 0.1 $ & $ 42.667 \pm 0.1 $ & $ 4.5 _ {-1.4} ^ {+1.0} $ & $ 4.3 _ {-0.6} ^ {+0.6} $ & \citep{2017ApJ...840...97F} \\
NGC4051 & 0.00308 & $ -10.327 \pm 0.11 $ & $ 41.969 \pm 0.11 $ & $ 5.7 _ {-1.4} ^ {+1.9} $ & $ 5.0 _ {-0.5} ^ {+0.5} $ & \citep{2017ApJ...840...97F} \\
3C120 & 0.03279 & $ -10.506 \pm 0.05 $ & $ 43.864 \pm 0.05 $ & $ 25.8 _ {-2.3} ^ {+2.0} $ & $ 29.5 _ {-0.9} ^ {+0.9} $ & \citep{2012ApJ...755...60G} \\
Mrk1501 & 0.08603 & $ -11.295 \pm 0.05 $ & $ 43.945 \pm 0.05 $ & $ 12.4 _ {-2.4} ^ {+2.2} $ & $ 14.4 _ {-0.7} ^ {+0.7} $ & \citep{2012ApJ...755...60G} \\
Mrk335 & 0.02461 & $ -10.477 \pm 0.06 $ & $ 43.639 \pm 0.06 $ & $ 17.2 _ {-1.8} ^ {+3.7} $ & $ 13.6 _ {-0.4} ^ {+0.4} $ & \citep{2012ApJ...755...60G} \\
Mrk6 & 0.01951 & $ -10.128 \pm 0.06 $ & $ 43.782 \pm 0.06 $ & $ 9.7 _ {-7.1} ^ {+3.6} $ & $ 8.5 _ {-0.5} ^ {+0.5} $ & \citep{2012ApJ...755...60G} \\
PG2130+099 & 0.06217 & $ -10.826 \pm 0.03 $ & $ 44.118 \pm 0.03 $ & $ 17.1 _ {-4.8} ^ {+2.4} $ & $ 11.3 _ {-0.6} ^ {+0.6} $ & \citep{2012ApJ...755...60G} \\
RM017 & 0.45657 & $ -12.972 \pm 0.01 $ & $ 43.891 \pm 0.01 $ & $ 21.8 _ {-6.9} ^ {+4.7} $ & $ 7.7 _ {-1.1} ^ {+2.1} $ & \citep{2017ApJ...851...21G} \\
RM191 & 0.44206 & $ -13.21 \pm 0.01 $ & $ 43.62 \pm 0.01 $ & $ 8.7 _ {-1.7} ^ {+1.8} $ & $ 9.0 _ {-0.9} ^ {+0.9} $ & \citep{2017ApJ...851...21G} \\
RM229 & 0.47008 & $ -13.364 \pm 0.01 $ & $ 43.53 \pm 0.01 $ & $ 14.3 _ {-4.1} ^ {+2.9} $ & $ 14.6 _ {-2.9} ^ {+2.3} $ & \citep{2017ApJ...851...21G} \\
RM265 & 0.7357 & $ -13.193 \pm 0.01 $ & $ 44.172 \pm 0.01 $ & $ 9.1 _ {-7.2} ^ {+4.6} $ & $ 7.6 _ {-2.5} ^ {+2.2} $ & \citep{2017ApJ...851...21G} \\
RM267 & 0.5877 & $ -13.056 \pm 0.01 $ & $ 44.071 \pm 0.01 $ & $ 20.3 _ {-2.5} ^ {+2.4} $ & $ 20.2 _ {-1.2} ^ {+1.1} $ & \citep{2017ApJ...851...21G} \\
RM272 & 0.26303 & $ -12.404 \pm 0.02 $ & $ 43.9 \pm 0.02 $ & $ 17.1 _ {-5.1} ^ {+4.9} $ & $ 13.5 _ {-1.9} ^ {+2.1} $ & \citep{2017ApJ...851...21G} \\
RM301 & 0.54846 & $ -13.024 \pm 0.01 $ & $ 44.031 \pm 0.01 $ & $ 11.9 _ {-4.7} ^ {+3.7} $ & $ 12.0 _ {-1.6} ^ {+1.6} $ & \citep{2017ApJ...851...21G} \\
RM320 & 0.26493 & $ -12.912 \pm 0.01 $ & $ 43.4 \pm 0.01 $ & $ 23.3 _ {-10.5} ^ {+5.8} $ & $ 24.0 _ {-2.4} ^ {+1.8} $ & \citep{2017ApJ...851...21G} \\
RM377 & 0.33701 & $ -13.213 \pm 0.01 $ & $ 43.34 \pm 0.01 $ & $ 7.1 _ {-2.9} ^ {+10.9} $ & $ 5.8 _ {-0.6} ^ {+0.6} $ & \citep{2017ApJ...851...21G} \\
RM392 & 0.8429 & $ -13.25 \pm 0.01 $ & $ 44.26 \pm 0.01 $ & $ 13.6 _ {-4.9} ^ {+3.2} $ & $ 12.8 _ {-2.1} ^ {+1.8} $ & \citep{2017ApJ...851...21G} \\
RM519 & 0.55421 & $ -13.915 \pm 0.01 $ & $ 43.15 \pm 0.01 $ & $ 11.0 _ {-3.8} ^ {+4.5} $ & $ 10.7 _ {-1.6} ^ {+1.6} $ & \citep{2017ApJ...851...21G} \\
RM551 & 0.68079 & $ -13.311 \pm 0.01 $ & $ 43.971 \pm 0.01 $ & $ 6.4 _ {-3.1} ^ {+7.4} $ & $ 6.5 _ {-1.6} ^ {+1.7} $ & \citep{2017ApJ...851...21G} \\
RM694 & 0.53277 & $ -12.893 \pm 0.01 $ & $ 44.132 \pm 0.01 $ & $ 14.9 _ {-5.5} ^ {+5.5} $ & $ 11.7 _ {-1.9} ^ {+2.0} $ & \citep{2017ApJ...851...21G} \\
RM775 & 0.17293 & $ -12.37 \pm 0.01 $ & $ 43.525 \pm 0.01 $ & $ 19.2 _ {-10.0} ^ {+6.6} $ & $ 13.9 _ {-3.4} ^ {+4.0} $ & \citep{2017ApJ...851...21G} \\
RM776 & 0.11655 & $ -12.428 \pm 0.01 $ & $ 43.094 \pm 0.01 $ & $ 10.1 _ {-2.9} ^ {+2.9} $ & $ 8.2 _ {-0.9} ^ {+1.0} $ & \citep{2017ApJ...851...21G} \\
RM779 & 0.15281 & $ -12.712 \pm 0.01 $ & $ 43.065 \pm 0.01 $ & $ 11.3 _ {-7.3} ^ {+5.0} $ & $ 10.2 _ {-1.3} ^ {+1.0} $ & \citep{2017ApJ...851...21G} \\
RM790 & 0.23754 & $ -12.972 \pm 0.01 $ & $ 43.232 \pm 0.01 $ & $ 8.8 _ {-2.8} ^ {+4.0} $ & $ 6.1 _ {-3.0} ^ {+5.4} $ & \citep{2017ApJ...851...21G} \\
RM840 & 0.2443 & $ -13.08 \pm 0.01 $ & $ 43.151 \pm 0.01 $ & $ 5.8 _ {-2.0} ^ {+2.1} $ & $ 6.7 _ {-1.5} ^ {+1.1} $ & \citep{2017ApJ...851...21G} \\
PG0026+129 & 0.14424 & $ -10.788 \pm 0.02 $ & $ 44.935 \pm 0.02 $ & $ 10.9 _ {-4.4} ^ {+5.3} $ & $ 15.2 _ {-3.4} ^ {+3.6} $ & \citep{2020ApJ...905...75H} \\
PG0026+129 & 0.14424 & $ -10.828 \pm 0.02 $ & $ 44.895 \pm 0.02 $ & $ 27.8 _ {-4.7} ^ {+3.7} $ & $ 25.6 _ {-2.5} ^ {+2.5} $ & \citep{2020ApJ...905...75H} \\
Mrk335 & 0.02461 & $ -10.727 \pm 0.05 $ & $ 43.389 \pm 0.05 $ & $ 9.3 _ {-2.1} ^ {+2.4} $ & $ 18.4 _ {-1.2} ^ {+1.3} $ & \citep{2021ApJS..253...20H} \\
PG0804+761 & 0.10006 & $ -10.46 \pm 0.04 $ & $ 44.921 \pm 0.04 $ & $ 69.7 _ {-7.2} ^ {+7.3} $ & $ 57.8 _ {-3.6} ^ {+3.7} $ & \citep{2021ApJS..253...20H} \\
PG1322+659 & 0.16769 & $ -11.157 \pm 0.03 $ & $ 44.709 \pm 0.03 $ & $ 31.8 _ {-4.5} ^ {+10.5} $ & $ 27.6 _ {-2.0} ^ {+2.1} $ & \citep{2021ApJS..253...20H} \\
PG1402+261 & 0.16404 & $ -11.107 \pm 0.05 $ & $ 44.738 \pm 0.05 $ & $ 65.8 _ {-5.6} ^ {+5.1} $ & $ 64.5 _ {-4.4} ^ {+4.6} $ & \citep{2021ApJS..253...20H} \\
PG1404+226 & 0.09908 & $ -11.243 \pm 0.03 $ & $ 44.128 \pm 0.03 $ & $ 20.0 _ {-3.9} ^ {+3.6} $ & $ 19.2 _ {-2.2} ^ {+2.9} $ & \citep{2021ApJS..253...20H} \\
PG1415+451 & 0.1142 & $ -11.265 \pm 0.04 $ & $ 44.238 \pm 0.04 $ & $ 30.5 _ {-4.1} ^ {+4.2} $ & $ 33.1 _ {-3.0} ^ {+3.0} $ & \citep{2021ApJS..253...20H} \\
PG1440+356 & 0.07958 & $ -10.819 \pm 0.05 $ & $ 44.35 \pm 0.05 $ & $ 33.2 _ {-9.5} ^ {+11.0} $ & $ 20.4 _ {-3.2} ^ {+3.4} $ & \citep{2021ApJS..253...20H} \\
PG1448+273 & 0.06506 & $ -10.739 \pm 0.03 $ & $ 44.246 \pm 0.03 $ & $ 31.4 _ {-4.4} ^ {+5.3} $ & $ 26.7 _ {-4.6} ^ {+5.0} $ & \citep{2021ApJS..253...20H} \\
PG1519+226 & 0.13656 & $ -11.171 \pm 0.06 $ & $ 44.5 \pm 0.06 $ & $ 70.8 _ {-6.7} ^ {+5.3} $ & $ 41.9 _ {-2.8} ^ {+3.8} $ & \citep{2021ApJS..253...20H} \\

\bottomrule\bottomrule
\end{tabular}
\end{threeparttable}%
}
\end{table*}
\end{turnpage}

\begin{turnpage}
\begin{table*}
\centering
\resizebox{2.7\columnwidth}{!}{%
\begin{threeparttable}
%\contcaption{}
\setlength{\tabcolsep}{18pt}
\begin{tabular}{lcccccr}
\toprule\toprule
Object & $z$ & log F$_{5100}$ ($\mathrm{erg \, s^{-1} \, cm^{-2}}$) & log L$_{5100}$ ($\mathrm{erg \, s^{-1}}$) & $\tau_{\mathrm{H\beta, ICCF}}$ (days) & $\tau_{\mathrm{H\beta, PyROA}}$ (days) & Reference \\
(1) & (2) & (3) & (4) & (5) & (6) & (7) \\
\midrule

PG1535+547 & 0.0391 & $ -10.663 \pm 0.02 $ & $ 43.864 \pm 0.02 $ & $ 25.9 _ {-3.0} ^ {+3.3} $ & $ 23.5 _ {-2.2} ^ {+2.2} $ & \citep{2021ApJS..253...20H} \\
PG1617+175 & 0.11482 & $ -11.019 \pm 0.02 $ & $ 44.49 \pm 0.02 $ & $ 32.5 _ {-2.3} ^ {+2.8} $ & $ 32.3 _ {-1.8} ^ {+1.9} $ & \citep{2021ApJS..253...20H} \\
PG1626+554 & 0.13361 & $ -11.026 \pm 0.06 $ & $ 44.624 \pm 0.06 $ & $ 78.2 _ {-3.2} ^ {+3.7} $ & $ 84.0 _ {-4.1} ^ {+4.1} $ & \citep{2021ApJS..253...20H} \\
PG0026+129 & 0.14424 & $ -10.798 \pm 0.02 $ & $ 44.925 \pm 0.02 $ & $ 109.8 _ {-29.7} ^ {+27.1} $ & $ 112.9 _ {-14.5} ^ {+12.5} $ & 
 \citep{2000ApJ...533..631K, 2004ApJ...613..682P} \\
PG0052+251 & 0.15339 & $ -11.041 \pm 0.03 $ & $ 44.74 \pm 0.03 $ & $ 90.0 _ {-24.6} ^ {+24.9} $ & $ 108.3 _ {-11.6} ^ {+9.9} $ & \citep{2000ApJ...533..631K, 2004ApJ...613..682P} \\
PG0804+761 & 0.10006 & $ -10.53 \pm 0.02 $ & $ 44.851 \pm 0.02 $ & $ 142.7 _ {-17.6} ^ {+20.2} $ & $ 143.4 _ {-43.7} ^ {+6.6} $ & \citep{2000ApJ...533..631K, 2004ApJ...613..682P} \\
PG0953+414 & 0.23487 & $ -11.06 \pm 0.01 $ & $ 45.133 \pm 0.01 $ & $ 153.4 _ {-19.4} ^ {+17.8} $ & $ 126.1 _ {-11.8} ^ {+13.1} $ & \citep{2000ApJ...533..631K, 2004ApJ...613..682P} \\
PG1226+023 & 0.15949 & $ -9.911 \pm 0.02 $ & $ 45.907 \pm 0.02 $ & $ 306.1 _ {-91.1} ^ {+70.9} $ & $ 123.9 _ {-14.4} ^ {+13.8} $ & \citep{2000ApJ...533..631K, 2004ApJ...613..682P} \\
PG1229+204 & 0.06458 & $ -11.316 \pm 0.06 $ & $ 43.662 \pm 0.06 $ & $ 36.8 _ {-15.7} ^ {+28.1} $ & $ 34.0 _ {-10.1} ^ {+13.8} $ & \citep{2000ApJ...533..631K, 2004ApJ...613..682P} \\
PG1617+175 & 0.11482 & $ -11.159 \pm 0.02 $ & $ 44.35 \pm 0.02 $ & $ 65.6 _ {-53.2} ^ {+33.1} $ & $ 49.8 _ {-21.2} ^ {+24.0} $ & \citep{2000ApJ...533..631K, 2004ApJ...613..682P} \\
PG0923+201 & 0.19368 & $ -10.747 \pm 0.06 $ & $ 45.258 \pm 0.06 $ & $ 92.3 _ {-10.9} ^ {+10.5} $ & $ 108.4 _ {-6.2} ^ {+5.8} $ & \citep{2021ApJ...922..142L} \\
PG1001+291 & 0.3301 & $ -10.973 \pm 0.03 $ & $ 45.559 \pm 0.03 $ & $ 42.8 _ {-7.3} ^ {+5.6} $ & $ 51.6 _ {-8.0} ^ {+2.4} $ & \citep{2021ApJ...922..142L} \\
NGC5548 & 0.01788 & $ -10.398 \pm 0.03 $ & $ 43.435 \pm 0.03 $ & $ 4.2 _ {-0.5} ^ {+0.5} $ & $ 4.0 _ {-0.3} ^ {+0.4} $ & \citep{2017ApJ...837..131P} \\
NGC5548 & 0.01788 & $ -10.358 \pm 0.09 $ & $ 43.475 \pm 0.09 $ & $ 21.5 _ {-2.4} ^ {+2.5} $ & $ 20.8 _ {-1.1} ^ {+1.2} $ & \citep{2002ApJ...581..197P, 2004ApJ...613..682P} \\
NGC5548 & 0.01788 & $ -10.478 \pm 0.09 $ & $ 43.355 \pm 0.09 $ & $ 16.6 _ {-1.3} ^ {+1.1} $ & $ 15.5 _ {-0.5} ^ {+0.5} $ & \citep{2002ApJ...581..197P, 2004ApJ...613..682P}  \\
NGC5548 & 0.01788 & $ -10.468 \pm 0.1 $ & $ 43.365 \pm 0.1 $ & $ 19.7 _ {-1.8} ^ {+1.5} $ & $ 20.3 _ {-0.5} ^ {+0.5} $ & \citep{2002ApJ...581..197P, 2004ApJ...613..682P}  \\
NGC5548 & 0.01788 & $ -10.508 \pm 0.1 $ & $ 43.325 \pm 0.1 $ & $ 13.8 _ {-3.9} ^ {+4.1} $ & $ 12.0 _ {-1.0} ^ {+1.1} $ & \citep{2002ApJ...581..197P, 2004ApJ...613..682P}  \\
NGC5548 & 0.01788 & $ -10.538 \pm 0.1 $ & $ 43.295 \pm 0.1 $ & $ 15.8 _ {-2.5} ^ {+3.1} $ & $ 16.4 _ {-0.9} ^ {+1.0} $ & \citep{2002ApJ...581..197P, 2004ApJ...613..682P} \\
NGC5548 & 0.01788 & $ -10.528 \pm 0.1 $ & $ 43.305 \pm 0.1 $ & $ 13.3 _ {-1.4} ^ {+1.2} $ & $ 13.9 _ {-0.7} ^ {+0.7} $ & \citep{2002ApJ...581..197P, 2004ApJ...613..682P}  \\
NGC5548 & 0.01788 & $ -10.728 \pm 0.11 $ & $ 43.105 \pm 0.11 $ & $ 18.6 _ {-2.3} ^ {+2.0} $ & $ 16.5 _ {-0.7} ^ {+0.7} $ & \citep{2002ApJ...581..197P, 2004ApJ...613..682P}  \\
NGC5548 & 0.01788 & $ -10.798 \pm 0.11 $ & $ 43.035 \pm 0.11 $ & $ 11.1 _ {-1.9} ^ {+1.9} $ & $ 12.5 _ {-0.8} ^ {+0.7} $ & \citep{2002ApJ...581..197P, 2004ApJ...613..682P}  \\
NGC7469 & 0.01503 & $ -10.313 \pm 0.05 $ & $ 43.368 \pm 0.05 $ & $ 16.7 _ {-2.5} ^ {+2.2} $ & $ 17.5 _ {-0.8} ^ {+0.9} $ & \citep{2014ApJ...795..149P} \\
3C120 & 0.03279 & $ -10.366 \pm 0.05 $ & $ 44.004 \pm 0.05 $ & $ 42.1 _ {-17.8} ^ {+30.4} $ & $ 108.1 _ {-1.7} ^ {+1.7} $ & \citep{1998PASP..110..660P, 2004ApJ...613..682P} \\
Ark120 & 0.03274 & $ -10.798 \pm 0.1 $ & $ 43.571 \pm 0.1 $ & $ 33.3 _ {-5.9} ^ {+3.2} $ & $ 29.6 _ {-4.6} ^ {+2.3} $ & \citep{1998PASP..110..660P, 2004ApJ...613..682P} \\
Mrk110 & 0.03584 & $ -10.965 \pm 0.05 $ & $ 43.484 \pm 0.05 $ & $ 27.2 _ {-6.7} ^ {+5.3} $ & $ 22.8 _ {-1.8} ^ {+1.7} $ & \citep{1998PASP..110..660P, 2004ApJ...613..682P} \\
Mrk335 & 0.02461 & $ -10.377 \pm 0.05 $ & $ 43.739 \pm 0.05 $ & $ 12.7 _ {-4.9} ^ {+5.8} $ & $ 12.7 _ {-2.3} ^ {+2.8} $ & \citep{1998PASP..110..660P, 2004ApJ...613..682P} \\
Mrk335 & 0.02461 & $ -10.457 \pm 0.06 $ & $ 43.659 \pm 0.06 $ & $ 16.6 _ {-4.1} ^ {+4.9} $ & $ 15.5 _ {-1.5} ^ {+1.6} $ & \citep{1998PASP..110..660P, 2004ApJ...613..682P} \\
Mrk509 & 0.03346 & $ -10.283 \pm 0.05 $ & $ 44.105 \pm 0.05 $ & $ 77.3 _ {-5.8} ^ {+6.0} $ & $ 86.7 _ {-3.3} ^ {+3.3} $ & \citep{1998PASP..110..660P, 2004ApJ...613..682P} \\
Mrk590 & 0.02553 & $ -10.587 \pm 0.06 $ & $ 43.561 \pm 0.06 $ & $ 29.0 _ {-4.3} ^ {+3.2} $ & $ 22.1 _ {-1.6} ^ {+1.8} $ & \citep{1998PASP..110..660P, 2004ApJ...613..682P} \\
Mrk590 & 0.02553 & $ -10.647 \pm 0.07 $ & $ 43.501 \pm 0.07 $ & $ 20.6 _ {-2.6} ^ {+3.7} $ & $ 20.3 _ {-1.8} ^ {+1.9} $ & \citep{1998PASP..110..660P, 2004ApJ...613..682P} \\
Mrk590 & 0.02553 & $ -10.857 \pm 0.08 $ & $ 43.291 \pm 0.08 $ & $ 28.5 _ {-4.6} ^ {+4.5} $ & $ 27.3 _ {-2.1} ^ {+2.3} $ & \citep{1998PASP..110..660P, 2004ApJ...613..682P} \\
Mrk590 & 0.02553 & $ -11.107 \pm 0.11 $ & $ 43.041 \pm 0.11 $ & $ 13.8 _ {-8.7} ^ {+7.1} $ & $ 11.1 _ {-7.1} ^ {+6.4} $ & \citep{1998PASP..110..660P, 2004ApJ...613..682P} \\
Mrk79 & 0.02258 & $ -10.354 \pm 0.07 $ & $ 43.685 \pm 0.07 $ & $ 16.1 _ {-7.4} ^ {+6.4} $ & $ 18.0 _ {-4.1} ^ {+4.1} $ & \citep{1998PASP..110..660P, 2004ApJ...613..682P} \\
Mrk79 & 0.02258 & $ -10.424 \pm 0.07 $ & $ 43.615 \pm 0.07 $ & $ 15.7 _ {-5.4} ^ {+6.6} $ & $ 20.3 _ {-5.2} ^ {+6.3} $ & \citep{1998PASP..110..660P, 2004ApJ...613..682P} \\
Mrk817 & 0.0316 & $ -10.603 \pm 0.05 $ & $ 43.734 \pm 0.05 $ & $ 19.1 _ {-3.9} ^ {+4.1} $ & $ 20.1 _ {-2.9} ^ {+2.9} $ & \citep{1998PASP..110..660P, 2004ApJ...613..682P} \\
Mrk817 & 0.0316 & $ -10.723 \pm 0.05 $ & $ 43.614 \pm 0.05 $ & $ 15.1 _ {-3.4} ^ {+3.8} $ & $ 17.1 _ {-1.2} ^ {+1.3} $ & \citep{1998PASP..110..660P, 2004ApJ...613..682P} \\
Mrk817 & 0.0316 & $ -10.723 \pm 0.05 $ & $ 43.614 \pm 0.05 $ & $ 32.7 _ {-7.2} ^ {+5.9} $ & $ 36.5 _ {-3.5} ^ {+3.3} $ & \citep{1998PASP..110..660P, 2004ApJ...613..682P} \\
3C390.3 & 0.0559 & $ -10.941 \pm 0.02 $ & $ 43.907 \pm 0.02 $ & $ 139.8 _ {-31.1} ^ {+64.5} $ & $ 81.2 _ {-6.2} ^ {+9.4} $ & \citep{Shapovalova} \\
NGC3783 & 0.01079 & $ -10.75 \pm 0.18 $ & $ 42.641 \pm 0.18 $ & $ 3.8 _ {-3.5} ^ {+3.0} $ & $ 6.4 _ {-1.6} ^ {+1.7} $ & \citep{1994ApJ...425..609S, 2004ApJ...613..682P} \\
3C382 & 0.05512 & $ -10.749 \pm 0.09 $ & $ 44.086 \pm 0.09 $ & $ 9.7 _ {-6.7} ^ {+6.8} $ & $ 6.7 _ {-3.3} ^ {+3.9} $ & \citep{2022ApJ...925...52U} \\

\bottomrule\bottomrule
\end{tabular}
\end{threeparttable}%
}
\end{table*}
\end{turnpage}

\begin{turnpage}
\begin{table*}
\centering
\resizebox{2.7\columnwidth}{!}{%
\begin{threeparttable}
%\contcaption{}
%\caption{Best sample}
\setlength{\tabcolsep}{18pt}
\begin{tabular}{lcccccr}
\toprule\toprule
Object & $z$ & log F$_{5100}$ ($\mathrm{erg \, s^{-1} \, cm^{-2}}$) & log L$_{5100}$ ($\mathrm{erg \, s^{-1}}$) & $\tau_{\mathrm{H\beta, ICCF}}$ (days) & $\tau_{\mathrm{H\beta, PyROA}}$ (days) & Reference \\
(1) & (2) & (3) & (4) & (5) & (6) & (7) \\
\midrule
Ark120 & 0.03274 & $ -10.458 \pm 0.16 $ & $ 43.911 \pm 0.16 $ & $ 19.3 _ {-5.5} ^ {+8.6} $ & $ 21.2 _ {-4.3} ^ {+4.8} $ & \citep{2022ApJ...925...52U} \\
MCG+04-22-042 & 0.03405 & $ -11.208 \pm 0.11 $ & $ 43.196 \pm 0.11 $ & $ 13.6 _ {-1.9} ^ {+3.1} $ & $ 9.9 _ {-1.2} ^ {+1.0} $ & \citep{2022ApJ...925...52U} \\
Mrk110 & 0.03584 & $ -10.625 \pm 0.1 $ & $ 43.824 \pm 0.1 $ & $ 27.5 _ {-4.9} ^ {+4.0} $ & $ 16.7 _ {-0.8} ^ {+1.2} $ & \citep{2022ApJ...925...52U} \\
Mrk1392 & 0.03652 & $ -11.306 \pm 0.16 $ & $ 43.16 \pm 0.16 $ & $ 28.1 _ {-4.3} ^ {+3.2} $ & $ 41.1 _ {-1.2} ^ {+1.6} $ & \citep{2022ApJ...925...52U} \\
Mrk704 & 0.02962 & $ -10.528 \pm 0.04 $ & $ 43.752 \pm 0.04 $ & $ 26.0 _ {-7.3} ^ {+8.8} $ & $ 34.9 _ {-3.7} ^ {+1.7} $ & \citep{2022ApJ...925...52U} \\
NPM1G+27.0587 & 0.05946 & $ -10.952 \pm 0.04 $ & $ 43.952 \pm 0.04 $ & $ 13.5 _ {-4.2} ^ {+3.8} $ & $ 5.8 _ {-1.0} ^ {+4.5} $ & \citep{2022ApJ...925...52U} \\
PG2209+184 & 0.06873 & $ -11.742 \pm 0.11 $ & $ 43.293 \pm 0.11 $ & $ 13.6 _ {-2.5} ^ {+2.8} $ & $ 16.5 _ {-1.2} ^ {+1.1} $ & \citep{2022ApJ...925...52U} \\
RBS1303 & 0.04277 & $ -11.256 \pm 0.07 $ & $ 43.351 \pm 0.07 $ & $ 18.8 _ {-4.0} ^ {+3.0} $ & $ 19.3 _ {-3.7} ^ {+2.7} $ & \citep{2022ApJ...925...52U} \\
RBS1917 & 0.06477 & $ -11.358 \pm 0.04 $ & $ 43.623 \pm 0.04 $ & $ 11.9 _ {-4.3} ^ {+4.2} $ & $ 10.6 _ {-1.0} ^ {+0.9} $ & \citep{2022ApJ...925...52U} \\
RXJ2044.0+2833 & 0.04907 & $ -11.077 \pm 0.04 $ & $ 43.653 \pm 0.04 $ & $ 14.1 _ {-1.8} ^ {+1.8} $ & $ 12.1 _ {-1.9} ^ {+2.0} $ & \citep{2022ApJ...925...52U} \\
Zw535-012 & 0.04673 & $ -11.064 \pm 0.06 $ & $ 43.623 \pm 0.06 $ & $ 19.2 _ {-4.0} ^ {+6.9} $ & $ 13.7 _ {-1.5} ^ {+1.8} $ & \citep{2022ApJ...925...52U} \\
PG1226+023 & 0.15949 & $ -9.951 \pm 0.05 $ & $ 45.867 \pm 0.05 $ & $ 156.4 _ {-8.4} ^ {+9.4} $ & $ 218.4 _ {-58.4} ^ {+44.9} $ & \citep{2019ApJ...876...49Z} \\
PG2130+099 & 0.06217 & $ -10.586 \pm 0.01 $ & $ 44.358 \pm 0.01 $ & $ 20.3 _ {-4.2} ^ {+2.9} $ & $ 23.6 _ {-1.9} ^ {+2.3} $ & \citep{2008ApJ...688..837G} \\
NGC5548 & 0.01788 & $ -11.008 \pm 0.06 $ & $ 42.825 \pm 0.06 $ & $ 7.5 _ {-1.0} ^ {+1.4} $ & $ 7.6 _ {-0.5} ^ {+0.6} $ & \citep{2016ApJ...827..118L} \\
Mrk1501 & 0.08603 & $ -11.182 \pm 0.02 $ & $ 44.059 \pm 0.02 $ & $ 10.8 _ {-8.2} ^ {+7.0} $ & $ 17.1 _ {-2.2} ^ {+2.0} $ & \citep{2024ApJ...962...67W} \\
PG0052+251 & 0.15339 & $ -11.035 \pm 0.01 $ & $ 44.746 \pm 0.01 $ & $ 55.4 _ {-9.5} ^ {+10.4} $ & $ 30.4 _ {-2.2} ^ {+2.1} $ & \citep{2024ApJ...962...67W} \\
J0101+422 & 0.18909 & $ -11.166 \pm 0.01 $ & $ 44.815 \pm 0.01 $ & $ 64.0 _ {-9.9} ^ {+11.1} $ & $ 51.8 _ {-6.8} ^ {+5.0} $ & \citep{2024ApJ...962...67W} \\
J0140+234 & 0.32204 & $ -11.37 \pm 0.01 $ & $ 45.136 \pm 0.01 $ & $ 85.9 _ {-7.7} ^ {+7.2} $ & $ 84.6 _ {-4.8} ^ {+5.7} $ & \citep{2024ApJ...962...67W} \\
Mrk1014 & 0.1622 & $ -10.969 \pm 0.02 $ & $ 44.865 \pm 0.02 $ & $ 92.6 _ {-20.8} ^ {+18.5} $ & $ 179.7 _ {-7.6} ^ {+4.9} $ & \citep{2024ApJ...962...67W} \\
J0939+375 & 0.23194 & $ -11.717 \pm 0.01 $ & $ 44.464 \pm 0.01 $ & $ 15.7 _ {-11.1} ^ {+7.4} $ & $ 14.9 _ {-3.2} ^ {+3.1} $ & \citep{2024ApJ...962...67W} \\
PG0947+396 & 0.20633 & $ -11.485 \pm 0.01 $ & $ 44.582 \pm 0.01 $ & $ 30.4 _ {-9.1} ^ {+7.9} $ & $ 42.7 _ {-2.5} ^ {+2.5} $ & \citep{2024ApJ...962...67W} \\
J1026+523 & 0.25979 & $ -11.809 \pm 0.01 $ & $ 44.483 \pm 0.01 $ & $ 27.1 _ {-3.2} ^ {+3.3} $ & $ 22.7 _ {-1.6} ^ {+1.6} $ & \citep{2024ApJ...962...67W} \\
PG1100+772 & 0.31163 & $ -10.835 \pm 0.01 $ & $ 45.639 \pm 0.01 $ & $ 40.9 _ {-17.4} ^ {+10.9} $ & $ 52.0 _ {-11.1} ^ {+11.7} $ & \citep{2024ApJ...962...67W} \\
J1120+423 & 0.22702 & $ -11.625 \pm 0.01 $ & $ 44.534 \pm 0.01 $ & $ 36.2 _ {-12.1} ^ {+12.7} $ & $ 21.7 _ {-5.4} ^ {+5.3} $ & \citep{2024ApJ...962...67W} \\
PG1121+422 & 0.22568 & $ -11.311 \pm 0.01 $ & $ 44.843 \pm 0.01 $ & $ 94.5 _ {-16.5} ^ {+19.7} $ & $ 89.0 _ {-8.5} ^ {+8.1} $ & \citep{2024ApJ...962...67W} \\
PG1202+281 & 0.166 & $ -11.411 \pm 0.01 $ & $ 44.446 \pm 0.01 $ & $ 33.0 _ {-7.3} ^ {+7.8} $ & $ 16.0 _ {-3.3} ^ {+3.1} $ & \citep{2024ApJ...962...67W} \\
J1217+333 & 0.17888 & $ -11.743 \pm 0.02 $ & $ 44.185 \pm 0.02 $ & $ 22.5 _ {-17.6} ^ {+18.0} $ & $ 23.2 _ {-4.3} ^ {+4.1} $ & \citep{2024ApJ...962...67W} \\
VIII-Zw218 & 0.12763 & $ -11.143 \pm 0.01 $ & $ 44.465 \pm 0.01 $ & $ 56.1 _ {-13.7} ^ {+14.5} $ & $ 29.9 _ {-5.4} ^ {+5.1} $ & \citep{2024ApJ...962...67W} \\
PG1322+659 & 0.16769 & $ -11.058 \pm 0.01 $ & $ 44.808 \pm 0.01 $ & $ 42.2 _ {-14.2} ^ {+16.4} $ & $ 42.7 _ {-8.5} ^ {+7.3} $ & \citep{2024ApJ...962...67W} \\
J1415+483 & 0.27495 & $ -11.709 \pm 0.01 $ & $ 44.64 \pm 0.01 $ & $ 19.8 _ {-8.7} ^ {+9.3} $ & $ 27.5 _ {-4.1} ^ {+4.5} $ & \citep{2024ApJ...962...67W} \\
PG1427+480 & 0.22105 & $ -11.329 \pm 0.01 $ & $ 44.805 \pm 0.01 $ & $ 27.3 _ {-15.9} ^ {+16.9} $ & $ 33.4 _ {-4.1} ^ {+4.8} $ & \citep{2024ApJ...962...67W} \\
PG1440+356 & 0.07958 & $ -10.602 \pm 0.01 $ & $ 44.567 \pm 0.01 $ & $ 46.9 _ {-19.4} ^ {+15.7} $ & $ 30.3 _ {-4.1} ^ {+4.3} $ & \citep{2024ApJ...962...67W} \\
J1456+380 & 0.28392 & $ -11.647 \pm 0.01 $ & $ 44.733 \pm 0.01 $ & $ 60.6 _ {-6.8} ^ {+7.3} $ & $ 37.2 _ {-5.1} ^ {+6.3} $ & \citep{2024ApJ...962...67W} \\
J1540+355 & 0.16379 & $ -11.395 \pm 0.01 $ & $ 44.449 \pm 0.01 $ & $ 49.8 _ {-12.5} ^ {+15.6} $ & $ 43.7 _ {-5.4} ^ {+7.3} $ & \citep{2024ApJ...962...67W} \\
PG1612+261 & 0.13116 & $ -10.922 \pm 0.02 $ & $ 44.711 \pm 0.02 $ & $ 56.3 _ {-12.6} ^ {+12.0} $ & $ 50.6 _ {-8.0} ^ {+20.4} $ & \citep{2024ApJ...962...67W} \\
J1619+501 & 0.23366 & $ -11.769 \pm 0.01 $ & $ 44.419 \pm 0.01 $ & $ 26.2 _ {-5.4} ^ {+5.5} $ & $ 18.8 _ {-2.4} ^ {+2.4} $ & \citep{2024ApJ...962...67W} \\
PG2349-014 & 0.17261 & $ -11.242 \pm 0.02 $ & $ 44.652 \pm 0.02 $ & $ 47.7 _ {-9.2} ^ {+9.4} $ & $ 49.5 _ {-2.8} ^ {+2.6} $ & \citep{2024ApJ...962...67W} \\

\bottomrule\bottomrule
\end{tabular}
\begin{tablenotes}
\item Columns are: (1) object name, (2) redshift corrected for peculiar velocity, (3)  flux at 5100 {\AA} in log scale, (4) luminosity at 5100 {\AA} in log scale, (5) rest-frame H$\beta$ time lag from ICCF, (6) rest-frame H$\beta$ time lag from {\tt PyROA}, and (7) References correspond to the light curve data and the initial H$\beta$ lag. For AGNs with two references listed, the first pertains to the light curve data and the second to the published initial H$\beta$ lag. Note that luminosities are derived for the flat $\Lambda$CDM model (with $H_0=72\,{\rm km\,s^{-1}\,Mpc^{-1}}$ and $\Omega_{m0}=0.3$) using peculiar velocity corrected redshifts.
\end{tablenotes}
\end{threeparttable}%
}
\end{table*}
\end{turnpage}

A uniform lag analysis was conducted on this sample using three different methods: the Interpolated Cross-Correlation Function \citep[ICCF;][]{1986ApJ...305..175G, 1987ApJS...65....1G}, {\tt PyROA} \citep{2021MNRAS.508.5449D}, and {\tt JAVELIN} \citep{2011ApJ...735...80Z}. Several lag quality criteria were applied to ensure the reliability of the re-measured H$\beta$ lags: $\tau_{\mathrm{ICCF}} - \sigma_{\mathrm{ICCF}} > 0$, $r_{\text{max}} > 0.55$, $p$-value $\leq 0.10$, and $f_{\text{peak}} \geq 0.6$, \citep{2024ApJS..275...13W}. Here $\sigma_{\mathrm{ICCF}}$ represents the $1\sigma$ uncertainty in the ICCF lag, $r_{\text{max}}$ is the maximum of the cross-correlation coefficient, while the $p$-value and $f_{\text{peak}}$ quantify the reliability of the measured lags, see \citet{2024ApJS..275...13W} for details. After applying these criteria and conducting a thorough visual inspection, the final sample$-$referred to as the `Best sample'$-$comprised 157 high-quality measurements.

Figure \ref{fig:dist} presents the distribution of luminosities $L_{5100}$ and the number as a function of redshift for the Best sample, shown in red. For comparison, the distribution for the H$\alpha$-H$\beta$ sample from \citet{Caoetal2025} is shown in black. The Best sample spans a broad redshift range ($z = 0.00308$ to 0.8429) with a 16\% percentile of $ 0.01788$, a median of $0.06458$, and an 84\% percentile of $0.23371$. It also covers a significantly wider dynamic range in luminosity ($41.73 < \log L_{5100} \, (\mathrm{erg \, s^{-1}}) < 45.9$) than previously-used cosmological samples based on the H$\beta$ emission line, with a median value of $\log L_{5100} \, (\mathrm{erg \, s^{-1}})=43.79$, a 16\% percentile of $\log L_{5100} \, (\mathrm{erg \, s^{-1}})=43.07$, and an $84\%$ percentile of $\log L_{5100} \, (\mathrm{erg \, s^{-1}})=44.67$. This larger dataset allows for a more comprehensive examination of the $R-L$ relation across different cosmological models. In this study, we focus on the Best sample of 157 H$\beta$ RM measurements and adopt H$\beta$ lags determined using the ICCF method, as ICCF provides the most robust lag estimates \citep{2024ApJS..275...13W, 2025ApJ...985...30M}. A list of the sample properties used in our study is presented in Table~\ref{tab:sam_best}.

Given the inclusion of many low-redshift sources we applied redshift corrections to account for the Milky Way’s peculiar velocity, which more significantly affect measurements at low redshifts. To perform these corrections we used the NED Velocity Correction Calculator, which incorporates corrections for Galactic rotation, the peculiar motion of the Milky Way within the Local Group, the Local Group's infall toward the center of the Local Supercluster, and motion relative to the cosmic microwave background (CMB) reference frame. These corrections were applied uniformly to all AGNs in our study, resulting in redshifts that more accurately reflect their cosmological values by incorporating the effects of peculiar velocities. The corrected redshifts are listed in Table \ref{tab:sam_best}. Note that the rest-frame H$\beta$ lags listed in the same table have also been adjusted based on these updated redshifts.

\begin{figure}
    \centering
    \includegraphics[width=\columnwidth]{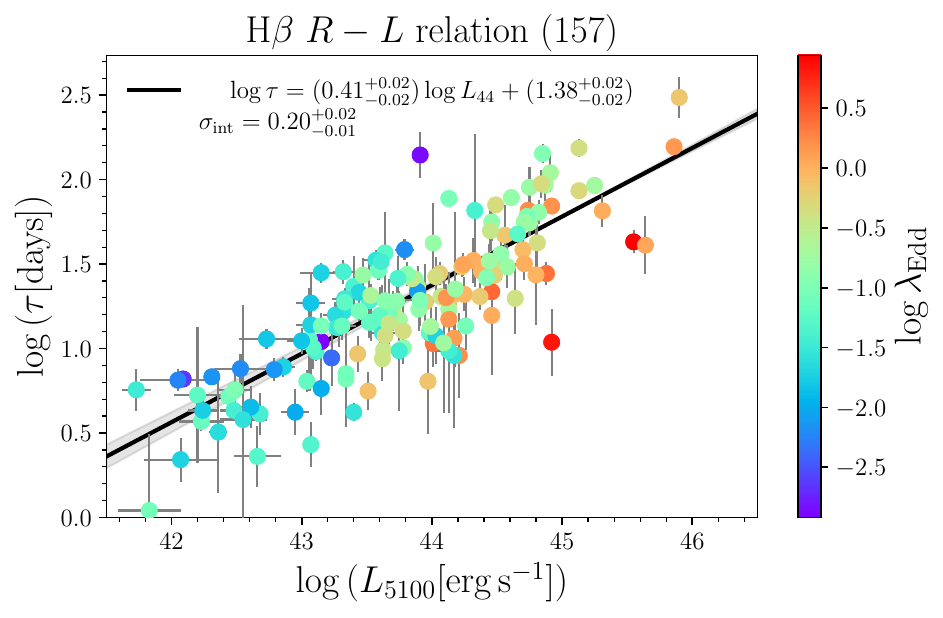}
    \caption{The radius$-$luminosity ($R-L$) relation for the Best H$\beta$ sample of 157 measurements evaluated in the fixed flat $\Lambda$CDM model (with $H_0=72\,{\rm km\,s^{-1}\,Mpc^{-1}}$ and $\Omega_{m0}=0.3$). The legend indicates the best-fit depicted by the black solid line ($L_{44}$ is the monochromatic luminosity at 5100\,\AA\, scaled to $10^{44}\,{\rm erg\,s^{-1}}$). The individual measurements are color-coded by the logarithm of the Eddington ratio $\lambda_{\rm Edd}$ with the color-scale to the right. The gray-shaded area around the best fit corresponds to the 1$\sigma$ uncertainty of the best fit. }
    \label{fig_RL_Edd_fixedLCDM}
\end{figure}

The Best sample exhibits a significant correlation between the rest-frame H$\beta$ time delay and the monochromatic luminosity at 5100\,\AA. In the fixed flat $\Lambda$CDM model with $H_0=72\,{\rm km\,s^{-1}\,Mpc^{-1}}$ and $\Omega_{m0}=0.3$, the Pearson correlation coefficient is $r=0.81$ ($p=2.29 \times 10^{-38}$) and the Spearman rank-order correlation coefficient is $s=0.78$ ($p=1.09 \times 10^{-33}$). Therefore, it is worth further studying the properties and consequences of the $R-L$ relation for this sample. The best-fit log-linear relation in the fixed flat $\Lambda$CDM model between the rest-frame time delay and the monochromatic luminosity is $\log{(\tau/\text{day})}=0.41^{+0.02}_{-0.02}\log{(L_{5100}/10^{44}\,{\rm erg\,s^{-1}})}+1.38^{+0.02}_{-0.02}$ with an intrinsic scatter of $\sigma_{\rm int}=0.20^{+0.02}_{-0.01}$ (here we apply the MCMC method using symmetrized time-delay uncertainties). In Fig.~\ref{fig_RL_Edd_fixedLCDM} we show the $R-L$ relation for the Best sample, where the individual measurements are color-coded with respect to the logarithm of the Eddington ratio ($\lambda_{\rm Edd}=L_{\rm bol}/L_{\rm Edd}$, see Subsection~\ref{sec:r_l} for further details). We see that the measurements at higher luminosities are also associated with higher Eddington ratios. We revisit the potential effects of the Eddington ratio (relative accretion rate) on the $R-L$ relation, especially its parameters and intrinsic scatter, in Subsection~\ref{sec:r_l}.

\subsection{$H(z)$ and BAO sample and data}

Additionally, we use 32 $H(z)$ measurements obtained from cosmic chronometers and 12 baryon acoustic oscillation (BAO) measurements for the analyses collectively referred as the $H(z)$ + BAO sample. This dataset covers a redshift range of $0.070 \leq z \leq 1.965$ for $H(z)$ and $0.122 \leq z \leq 2.334$ for BAO. Detailed descriptions of these data are provided in Tables 1 and 2, as well as in Sec. III, of \citet{CaoRatra2023}.

\section{Data Analysis Methodology}
\label{sec:analysis}

The correlation between the H$\beta$ rest-frame time lag ($\tau_{\mathrm{H}\beta}=R_{\mathrm{H}\beta}/c$) in days and the monochromatic luminosity at 5100\ \AA\ ($L_{5100}$) in units of $\rm erg\ s^{-1}$ can be expressed as
\begin{equation}
    \label{eq:R-L}
    \log{\left(\frac{\tau_{\mathrm{H}\beta}}{\rm day}\right)}=\beta+\gamma \log{\left(\frac{L_{5100}}{10^{44}\,{\rm erg\ s^{-1}}}\right)},
\end{equation}
where $\beta$ and $\gamma$ are the intercept and slope parameters, respectively (this is what we referred to as the H$\beta$ mono correlation in Ref.\ \citep{Caoetal2025}, where we studied a smaller sample of 41 AGN measurements). The luminosity $L_{5100}$ is derived from the measured AGN flux $F_{5100}$ (in $\rm erg\ s^{-1}\ cm^{-2}$) and the luminosity distance $D_L(z)$ is computed from Eq.\ \eqref{eq:DL}
\be
\label{eq:L5100}
    L_{5100}=4\pi D_L^2F_{5100}.
\ee

The natural logarithm of the likelihood function \citep{D'Agostini_2005} for H$\beta$ AGN data is given by 
\be
\label{eq:LH}
    \ln\mathcal{L}= -\frac{1}{2}\Bigg[\chi^2+\sum^{N}_{i=1}\ln\left(2\pi\sigma^2_{\mathrm{tot},i}\right)\Bigg],
\ee
where
\be
\label{eq:chi2}
    \chi^2 = \sum^{N}_{i=1}\bigg[\frac{(\log \tau_{\mathrm{H}\beta,i} - \beta - \gamma\log L_{5100,i})^2}{\sigma^2_{\mathrm{tot},i}}\bigg].
\ee
with the total uncertainty
\be
\label{eq:sigma_Hbeta}
\sigma^2_{\mathrm{tot},i}=\sigma_{\rm int}^2+\sigma_{{\log \tau_{\mathrm{H}\beta,i}}}^2+\gamma^2\sigma_{\log F_{5100,i}}^2.
\ee
Here $\sigma_{\rm int}$ denotes the intrinsic scatter associated with H$\beta$ AGNs, potentially encompassing unquantified systematic errors. The terms $\sigma_{{\log \tau_{\mathrm{H}\beta,i}}}$ and $\sigma_{{\log F_{5100,i}}}$ represent the uncertainties in the logarithm of the measured time delay and flux density for each of the $N$ data points. To handle the asymmetric errors in $\tau_{\mathrm{H}\beta}$,  we incorporate the upper error ($\sigma_{\tau_{\mathrm{H}\beta},+}$) as the $\tau_{\mathrm{H}\beta}$ error if the theoretical prediction for $\log{\tau_{\mathrm{H}\beta}}$ is greater than the observed value, and the lower error ($\sigma_{\tau_{\mathrm{H}\beta},-}$) otherwise.

Details regarding the likelihood functions for the $H(z)$ and BAO data, along with the description of their covariance matrices, can be found in Secs.\ IV and III, respectively, of Ref.\ \cite{CaoRatra2023}. 

For assessing the relative fit of different cosmological models to these data, we employ the Akaike information criterion (AIC), the Bayesian information criterion (BIC), and the more robust deviance information criterion (DIC). A detailed explanation of these criteria is provided in Sec.\ IV of Ref.\ \cite{CaoRatra2023}.

The Bayesian inference analyses are performed using the MCMC code {\footnotesize MontePython} \cite{Brinckmann2019}, assuming flat (uniform) prior probability distributions for all free parameters, as listed in Table \ref{tab:priors}. The resulting posterior statistics and visualizations are generated using the GetDist package \cite{Lewis_2019}.

\begin{table}[htbp]
\centering
% \resizebox{\columnwidth}{!}{%
\setlength\tabcolsep{3.3pt}
\begin{threeparttable}
\caption{Flat (uniform) priors of the constrained parameters.}
\label{tab:priors}
\begin{tabular}{lcc}
\toprule\toprule
Parameter & & Prior\\
\midrule
 & Cosmological parameters & \\
\midrule
$H_0$\,\tnote{a} &  & [None, None]\\
\obhs\,\tnote{b} &  & [0, 1]\\
\ochs\,\tnote{b} &  & [0, 1]\\
\ok &  & [$-2$, 2]\\
$\alpha$ &  & [0, 10]\\
\wx &  & [$-5$, 0.33]\\
\om\,\tnote{c} &  & [0.051314766115, 1]\\
% \midrule
\\
\midrule
 & $R-L$ relation parameters & \\
\midrule
$\beta$ &  & [0, 10]\\
$\gamma$ &  & [0, 5]\\
$\sigma_{\mathrm{int}}$ &  & [0, 5]\\
\bottomrule\bottomrule
\end{tabular}
\begin{tablenotes}
\item [a] \hunit. In the RM AGNs only cases, $H_0=70$ \hunit.
\item [b] Analyses involving $H(z)$ + BAO data. In the RM AGNs only cases $\Omega_{b}=0.05$.
\item [c] RM AGNs only, to ensure that $\Omega_{c}$ remains positive.
\end{tablenotes}
\end{threeparttable}%
% }
\end{table}

\section{Results}
\label{sec:results}

\begin{figure*}[htbp]
\centering
 \subfloat[Flat \lcdm]{%
    \includegraphics[width=0.45\textwidth,height=0.35\textwidth]{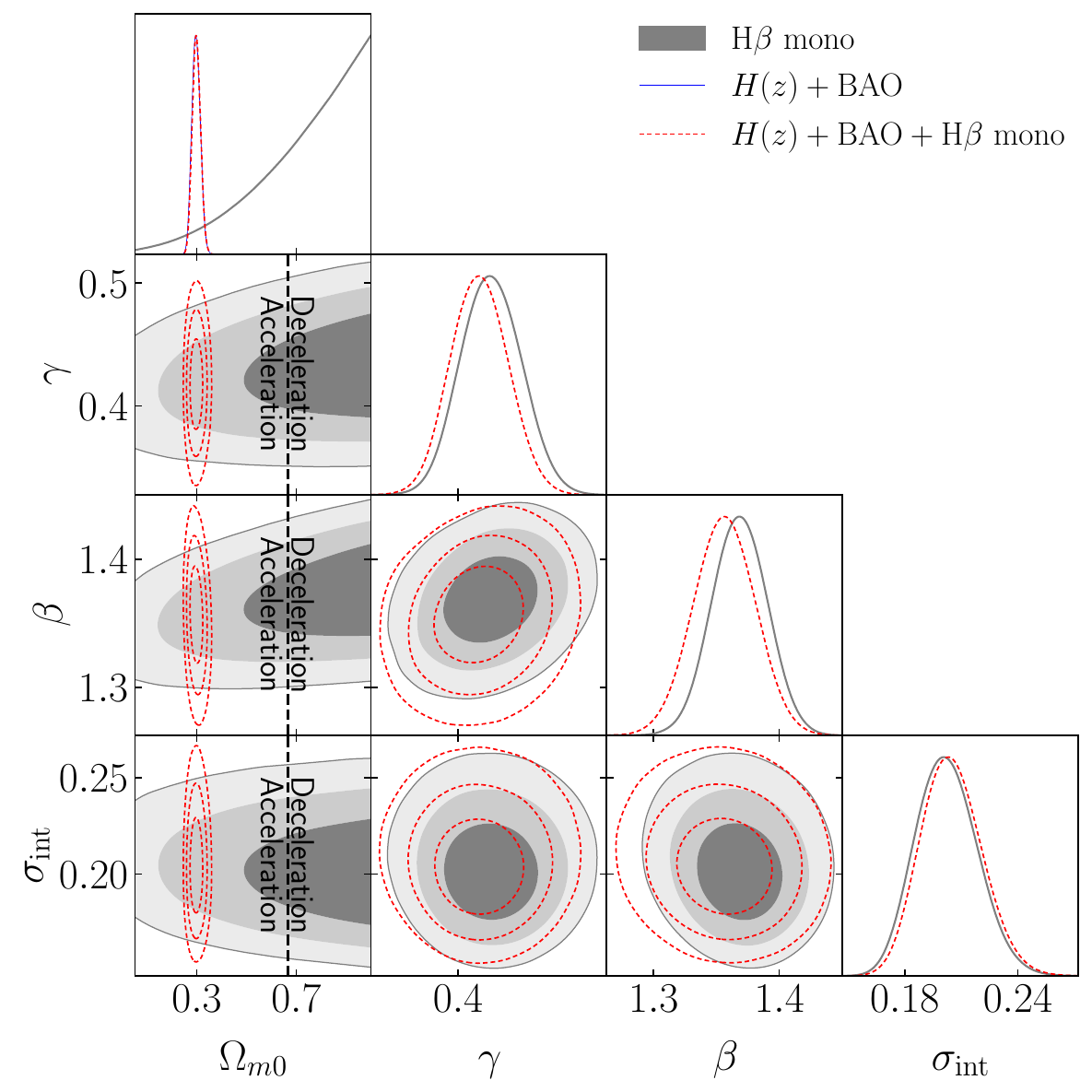}}
 \subfloat[Nonflat \lcdm]{%
    \includegraphics[width=0.45\textwidth,height=0.35\textwidth]{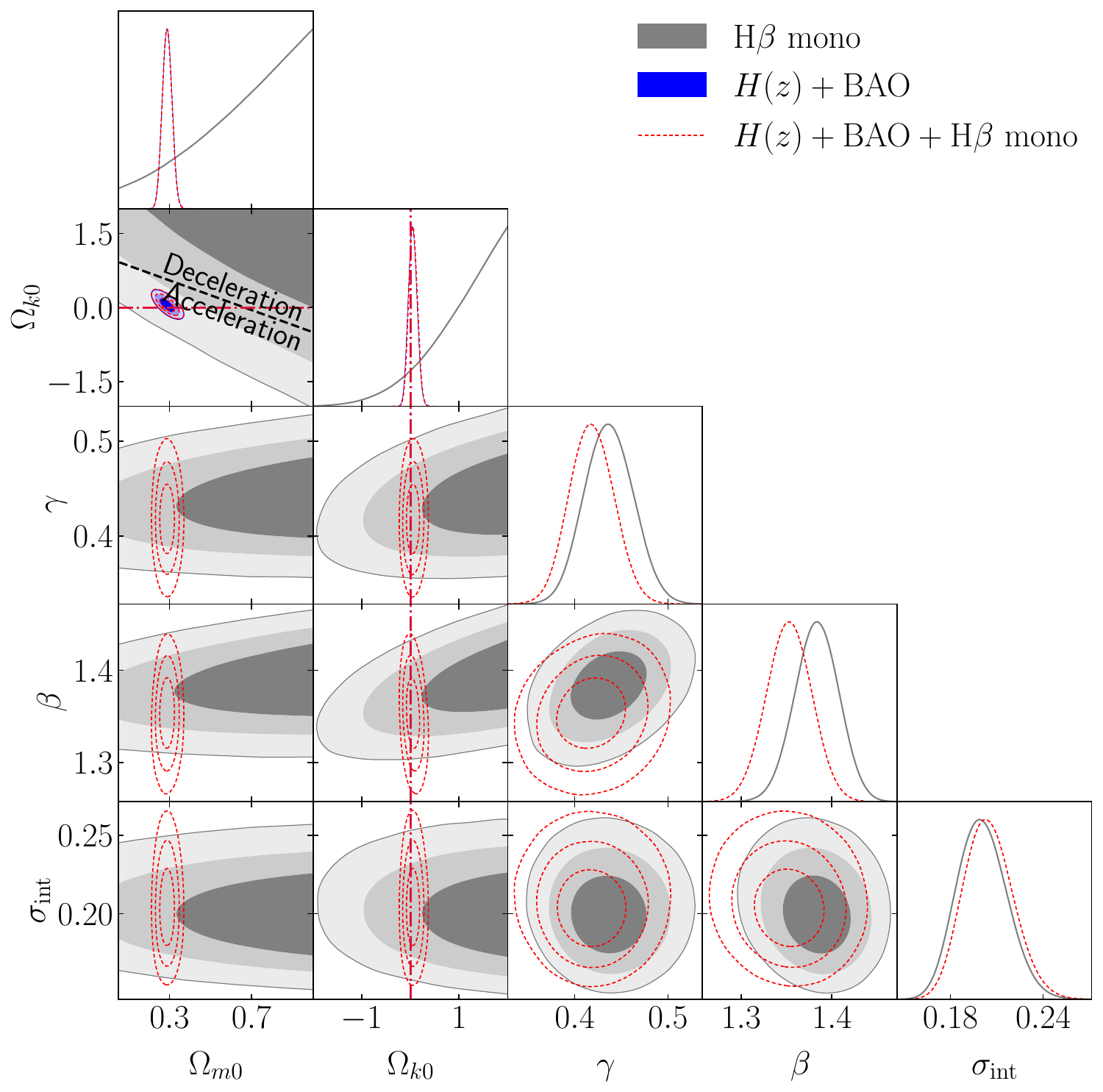}}\\
 \subfloat[Flat XCDM]{%
    \includegraphics[width=0.45\textwidth,height=0.35\textwidth]{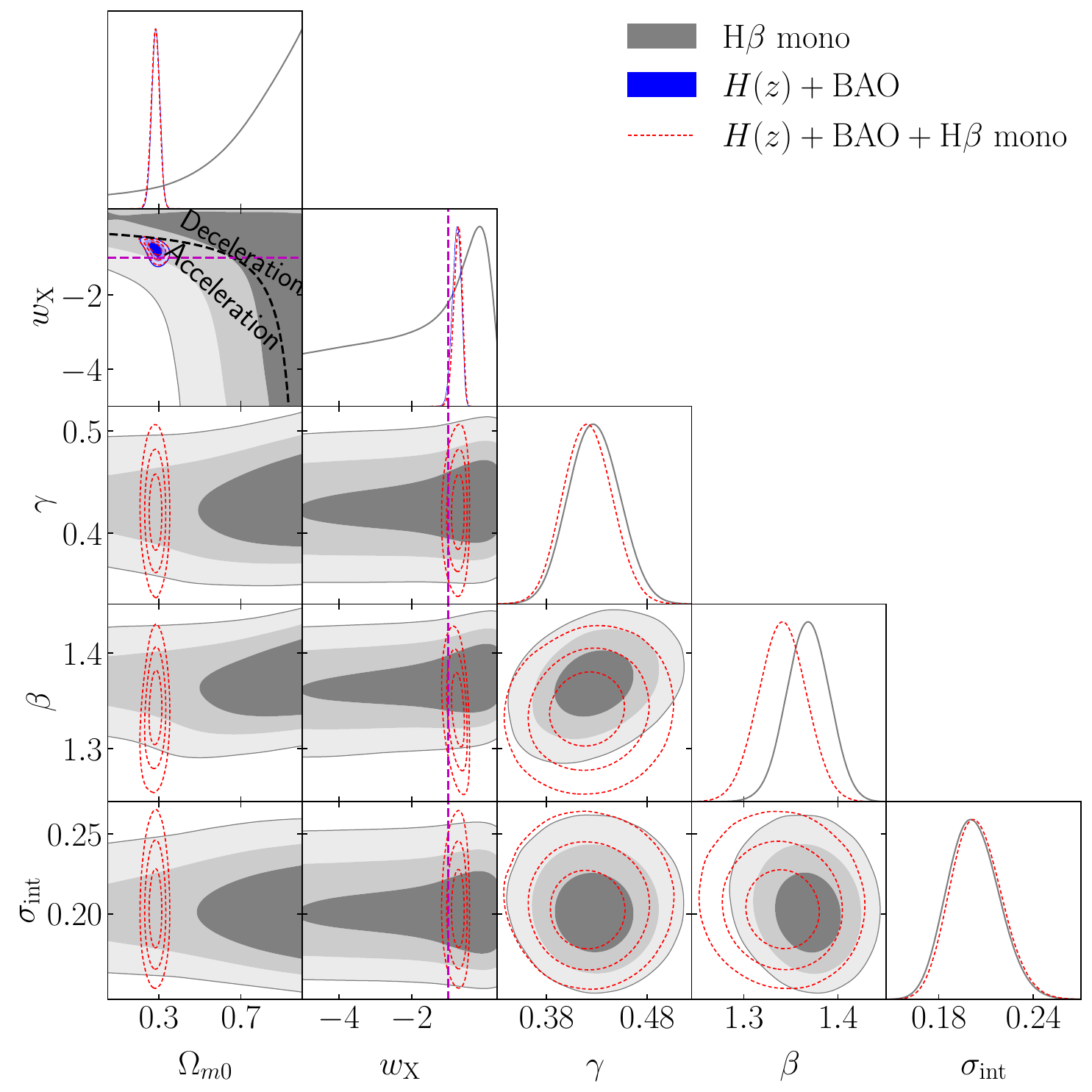}}
 \subfloat[Nonflat XCDM]{%
    \includegraphics[width=0.45\textwidth,height=0.35\textwidth]{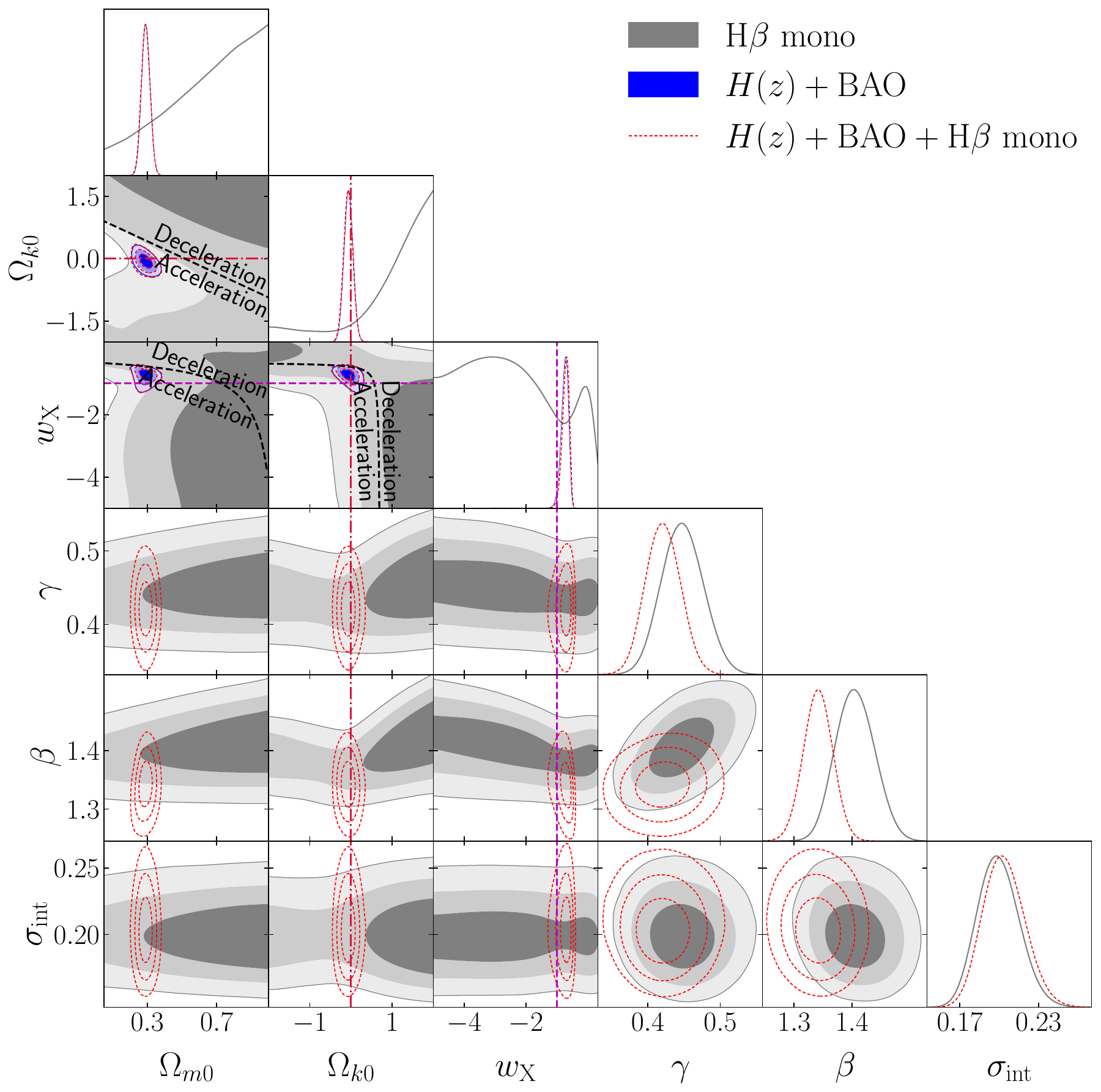}}\\
 \subfloat[Flat \pcdm]{%
    \includegraphics[width=0.45\textwidth,height=0.35\textwidth]{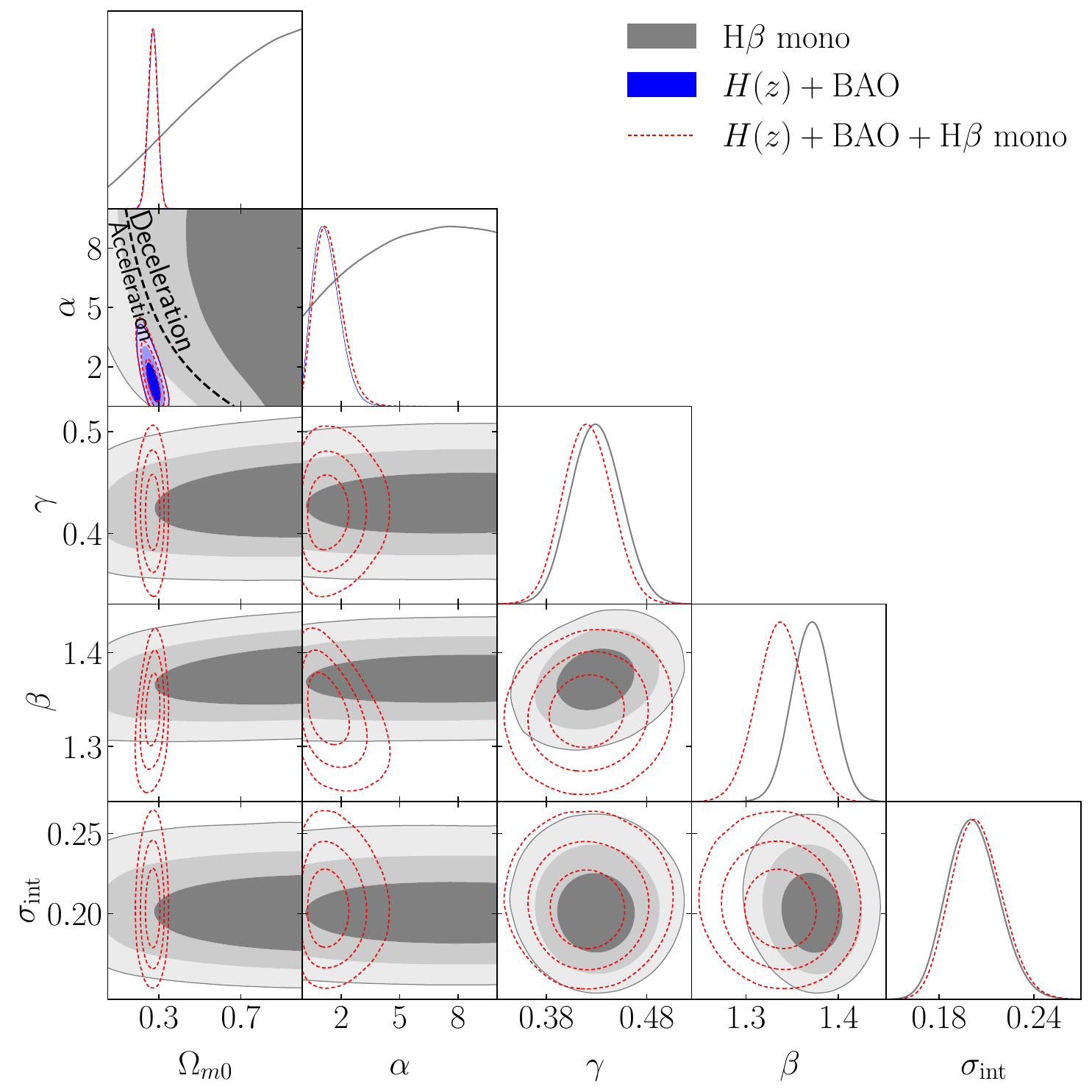}}
 \subfloat[Nonflat \pcdm]{%
    \includegraphics[width=0.45\textwidth,height=0.35\textwidth]{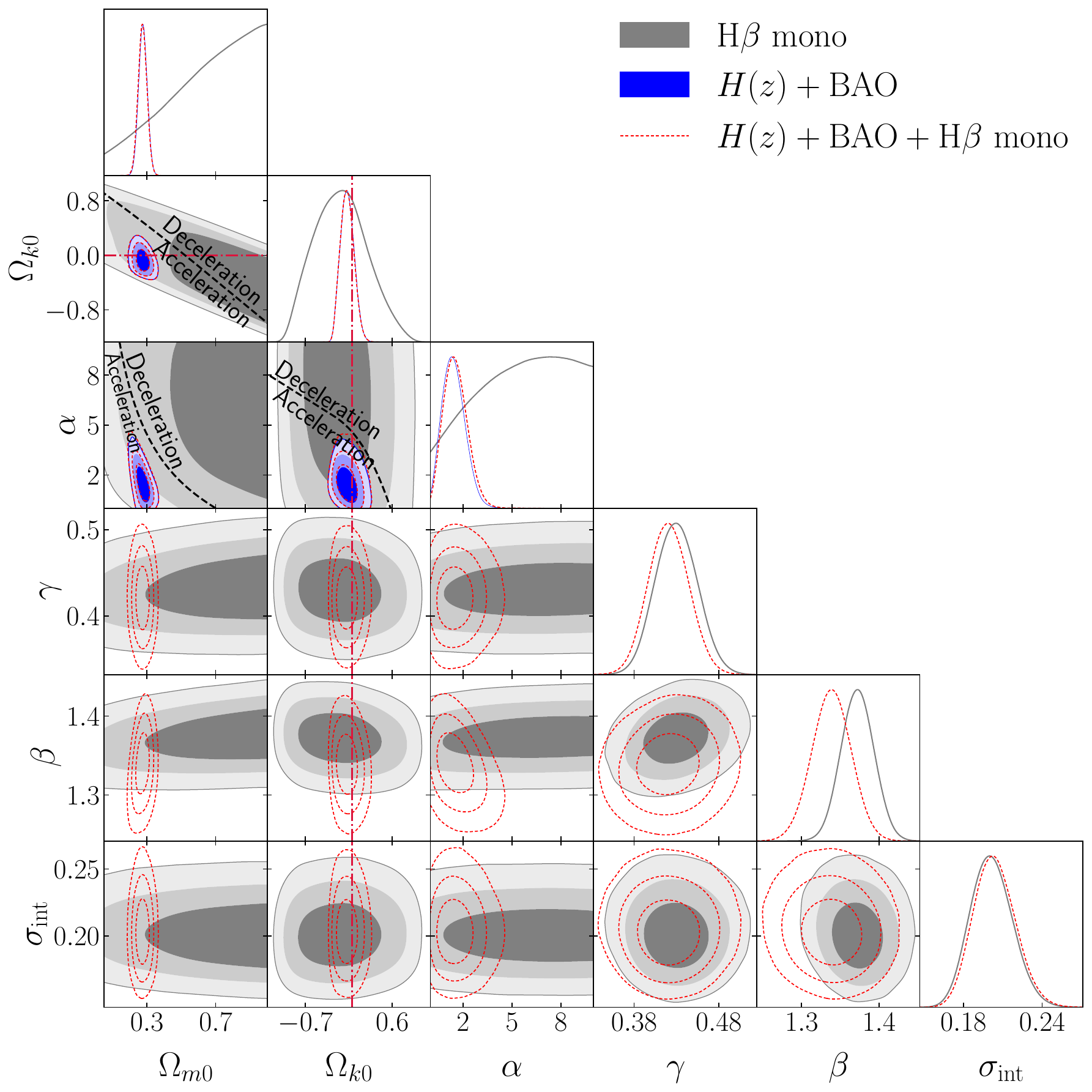}}\\
\caption{One-dimensional likelihoods and 1$\sigma$, 2$\sigma$, and 3$\sigma$ two-dimensional likelihood confidence contours from H$\beta$ mono (gray), $H(z)$ + BAO (blue), and $H(z)$ + BAO + H$\beta$ mono (dashed red) data for six different models, with \lcdm, XCDM, and \pcdm\ in the top, middle, and bottom rows, and flat (nonflat) models in the left (right) column. The black dashed zero-acceleration lines, computed for the third cosmological parameter set to the $H(z)$ + BAO data best-fitting values listed in Table \ref{tab:BFP} in panels (d) and (f), divide the parameter space into regions associated with currently-accelerating (below or below left) and currently-decelerating (above or above right) cosmological expansion. The crimson dash-dot lines represent flat hypersurfaces, with closed spatial hypersurfaces either below or to the left. The magenta lines represent $w_{\rm X}=-1$, i.e.\ flat or nonflat \lcdm\ models. The $\alpha = 0$ axes correspond to flat and nonflat \lcdm\ models in panels (e) and (f), respectively.}
\label{fig1}
\vspace{-50pt}
\end{figure*}

\begin{figure*}
\centering
 \subfloat[Flat \lcdm]{%
    \includegraphics[width=0.4\textwidth,height=0.35\textwidth]{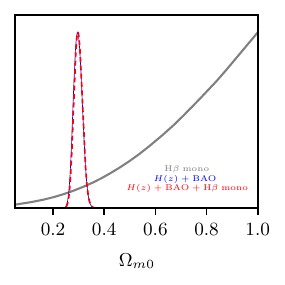}}
 \subfloat[Nonflat \lcdm]{%
    \includegraphics[width=0.4\textwidth,height=0.35\textwidth]{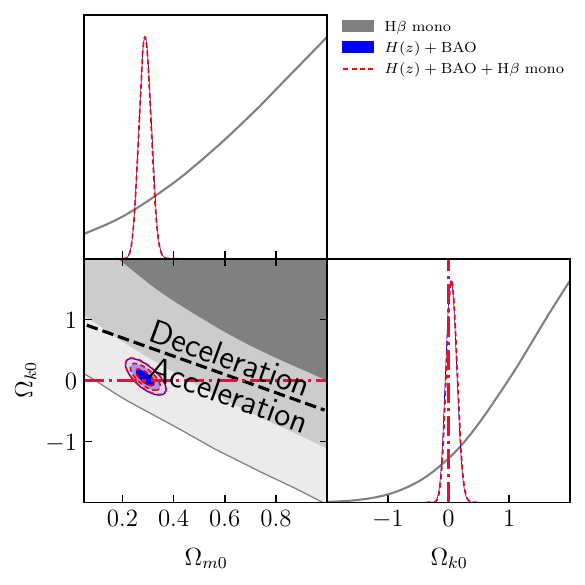}}\\
 \subfloat[Flat XCDM]{%
    \includegraphics[width=0.4\textwidth,height=0.35\textwidth]{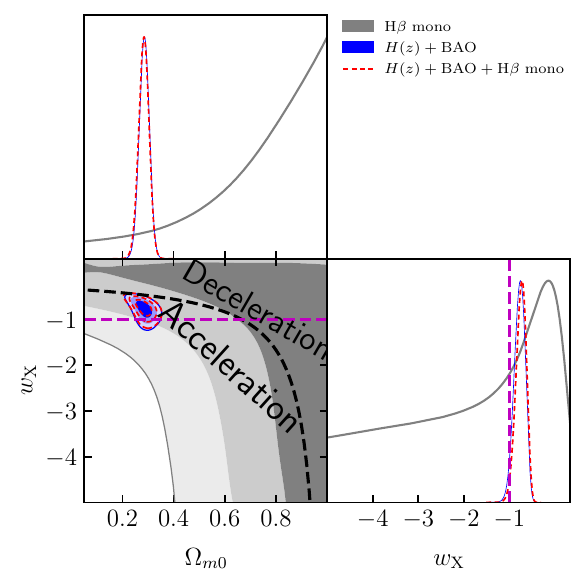}}
 \subfloat[Nonflat XCDM]{%
    \includegraphics[width=0.4\textwidth,height=0.35\textwidth]{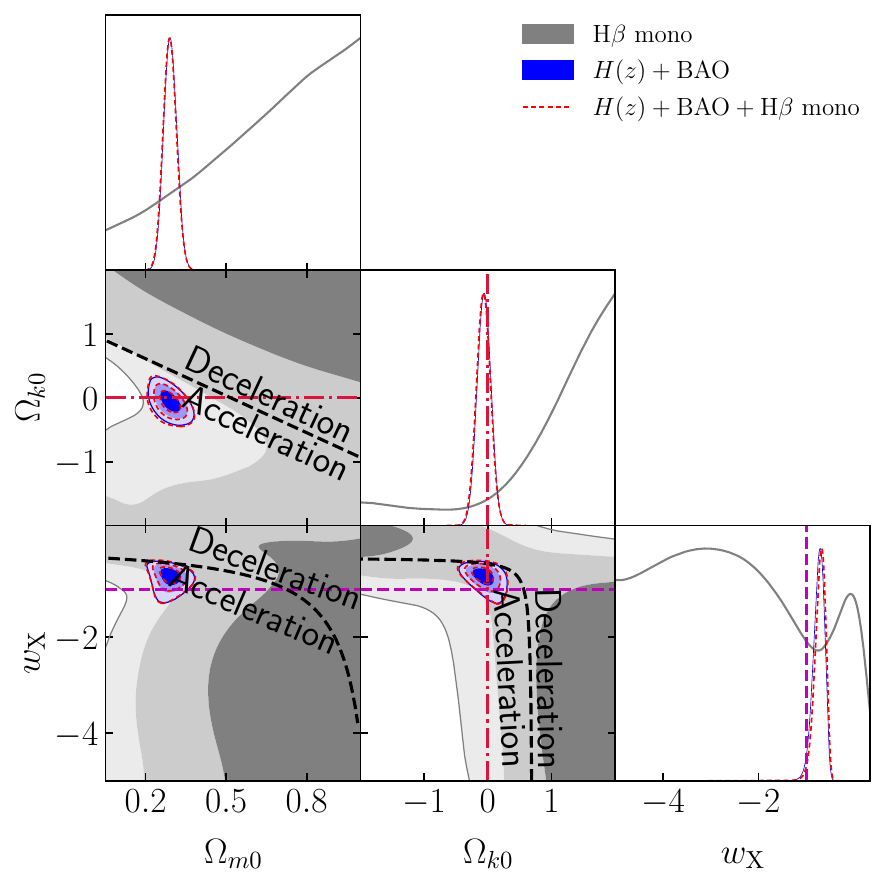}}\\
 \subfloat[Flat \pcdm]{%
    \includegraphics[width=0.4\textwidth,height=0.35\textwidth]{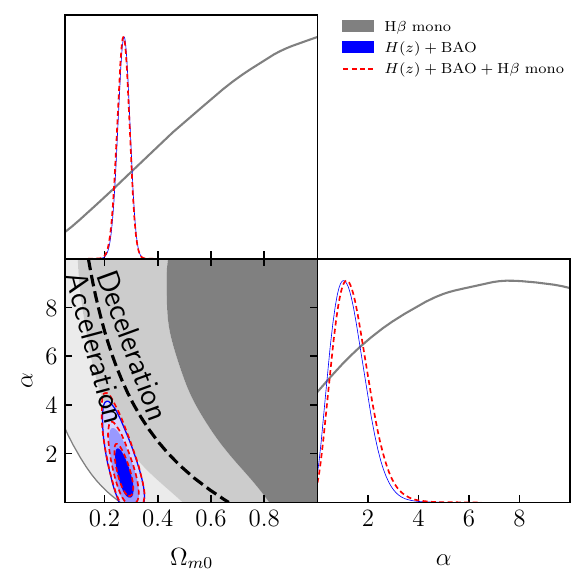}}
 \subfloat[Nonflat \pcdm]{%
    \includegraphics[width=0.4\textwidth,height=0.35\textwidth]{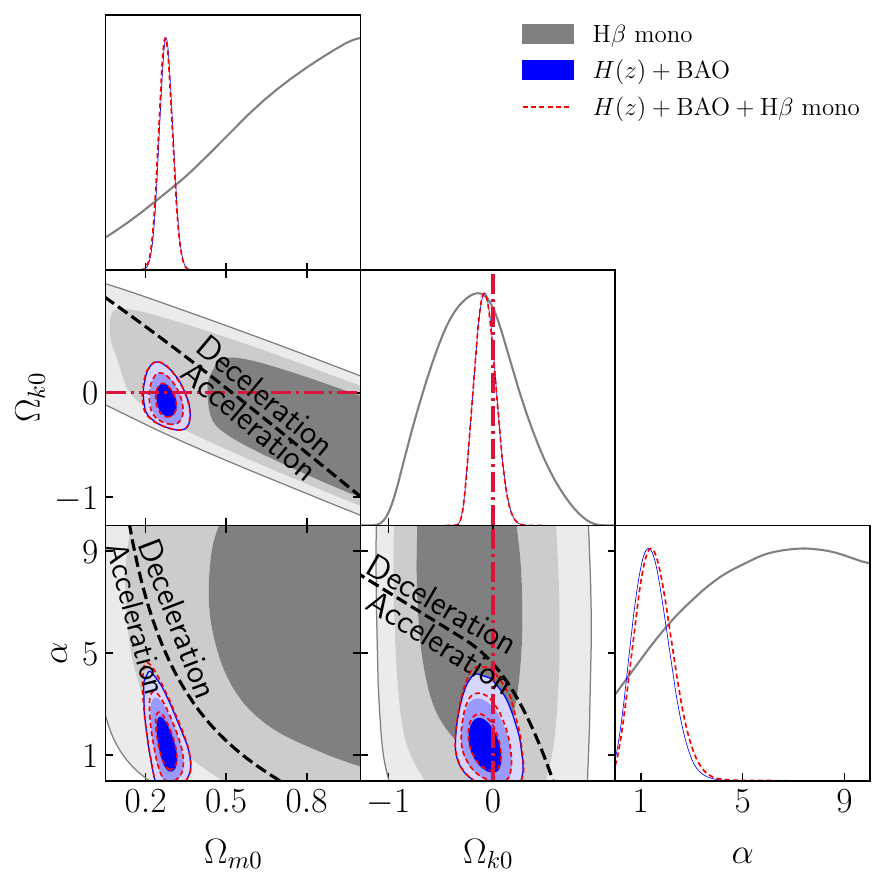}}\\
\caption{Same as Fig.\ \ref{fig1}, but for cosmological parameters only.}
\label{fig2}
\end{figure*}

Figures~\ref{fig1} and \ref{fig2} present the one-dimensional (1D) posterior probability distributions and 2D confidence regions for the cosmological parameters of the six cosmological models, alongside the H$\beta$ mono ($R_{\text{H}\beta} - L_{5100}$) relation and intrinsic scatter parameters. The 1D marginalized densities are normalized to a maximum of one. Tables \ref{tab:BFP} and \ref{tab:1d_BFP} summarize the unmarginalized best-fit parameters (with maximum likelihood $\mathcal{L}_{\rm max}$ and model selection criteria AIC, BIC, DIC, and their differences) and the marginalized posterior means with uncertainties, respectively.

\begin{turnpage}
\begin{table*}
\centering
% \footnotesize
% \fontsize{8.5pt}{10pt}\selectfont
% \resizebox*{2.7\columnwidth}{1.2\columnwidth}{%
\resizebox{2.7\columnwidth}{!}{%
\begin{threeparttable}
\caption{Unmarginalized best-fitting parameter values for all models from various combinations of data, where H$\beta$ mono stands for H$\beta$ time delays in combination with the monochromatic luminosity at 5100 \AA.}\label{tab:BFP}
\begin{tabular}{lccccccccccccccccc}
\toprule\toprule
Model & dataset & $\Omega_{b}h^2$ & $\Omega_{c}h^2$ & $\Omega_{m0}$ & $\Omega_{k0}$ & $w_{\mathrm{X}}$/$\alpha$\tnote{a} & $H_0$\tnote{b} & $\gamma$ & $\beta$ & $\sigma_{\mathrm{int}}$ & $-2\ln\mathcal{L}_{\mathrm{max}}$ & AIC & BIC & DIC & $\Delta \mathrm{AIC}$ & $\Delta \mathrm{BIC}$ & $\Delta \mathrm{DIC}$ \\
\midrule
 & $H(z)$ + BAO & 0.0254 & 0.1200 & 0.297 & $\cdots$ & $\cdots$ & 70.12 & $\cdots$ & $\cdots$ & $\cdots$ & 30.56 & 36.56 & 41.91 & 37.32 & 0.00 & 0.00 & 0.00\\%36.5576, 41.91016890175479, 37.318529778549525
Flat \lcdm & H$\beta$ mono & $\cdots$ & $\cdots$ & 0.999 & $\cdots$ & $\cdots$ & $\cdots$ & 0.433 & 1.373 & 0.198 & $-10.91$ & $-4.91$ & 4.26 & $-1.00$ & 0.00 & 0.00 & 0.00\\%-4.90816, 4.260577416044923, -0.9968786340207529
 & $H(z)$ + BAO + H$\beta$ mono & 0.0251 & 0.1210 & 0.300 & $\cdots$ & $\cdots$ & 69.99 & 0.418 & 1.355 & 0.203 & 24.64 & 36.64 & 56.46 & 37.54 & 0.00 & 0.00 & 0.00\\%36.6384, 56.45822944835446, 37.540251104032635
[6pt]
 & $H(z)$ + BAO & 0.0269 & 0.1128 & 0.289 & 0.041 & $\cdots$ & 69.61 & $\cdots$ & $\cdots$ & $\cdots$ & 30.34 & 38.34 & 45.48 & 38.80 & 1.78 & 3.56 & 1.48\\%38.3384, 45.475158535673046, 38.79768870799054
Nonflat \lcdm & H$\beta$ mono & $\cdots$ & $\cdots$ & 0.998 & 1.992 & $\cdots$ & $\cdots$ & 0.450 & 1.399 & 0.197 & $-13.51$ & $-5.51$ & 6.71 & $-2.25$ & $-0.60$ & 2.45 & $-1.25$\\%-5.5124, 6.712583221393231, -2.251784563516498
 & $H(z)$ + BAO + H$\beta$ mono & 0.0269 & 0.1131 & 0.291 & 0.041 & $\cdots$ & 69.58 & 0.420 & 1.355 & 0.200 & 24.30 & 38.30 & 61.42 & 39.05 & 1.66 & 4.96 & 1.51\\%38.2992, 61.42233435641353, 39.04661469652728
[6pt]
 & $H(z)$ + BAO & 0.0320 & 0.0932 & 0.283 & $\cdots$ & $-0.731$ & 66.69 & $\cdots$ & $\cdots$ & $\cdots$ & 26.57 & 34.57 & 41.71 & 34.52 & $-1.98$ & $-0.20$ & $-2.80$\\%34.5736, 41.710358535673045, 34.51703264021917
Flat XCDM & H$\beta$ mono & $\cdots$ & $\cdots$ & 0.066 & $\cdots$ & 0.141 & $\cdots$ & 0.427 & 1.378 & 0.197 & $-11.76$ & $-3.76$ & 8.46 & $-0.90$ & 1.15 & 4.20 & 0.09\\%-3.76208, 8.462903221393232, -0.9045998208701995
 & $H(z)$ + BAO + H$\beta$ mono & 0.0335 & 0.0886 & 0.278 & $\cdots$ & $-0.702$ & 66.48 & 0.420 & 1.342 & 0.197 & 19.47 & 33.47 & 56.60 & 33.63 & $-3.17$ & 0.14 & $-3.91$\\%33.47332, 56.59645435641353, 33.63396275105865
[6pt]
 & $H(z)$ + BAO & 0.0312 & 0.0990 & 0.293 & $-0.085$ & $-0.693$ & 66.84 & $\cdots$ & $\cdots$ & $\cdots$ & 26.00 & 36.00 & 44.92 & 36.17 & $-0.56$ & 3.01 & $-1.15$\\%35.996, 44.91694816959131, 36.17275325321086
Nonflat XCDM & H$\beta$ mono & $\cdots$ & $\cdots$ & 0.475 & $-1.974$ & 0.132 & $\cdots$ & 0.456 & 1.408 & 0.194 & $-15.28$ & $-5.28$ & 10.00 & $-3.25$ & $-0.37$ & 5.74 & $-2.25$\\%-5.28096, 10.000269026741538, -3.2481949113460367
 & $H(z)$ + BAO + H$\beta$ mono & 0.0329 & 0.0869 & 0.279 & $-0.062$ & $-0.656$ & 65.66 & 0.418 & 1.339 & 0.199 & 18.94 & 34.94 & 61.37 & 34.72 & $-1.69$ & 4.91 & $-2.82$\\%34.94342, 61.36985926447261, 34.716118404097486
[6pt]
 & $H(z)$ + BAO & 0.0337 & 0.0864 & 0.271 & $\cdots$ & 1.169 & 66.78 & $\cdots$ & $\cdots$ & $\cdots$ & 26.50 & 34.50 & 41.64 & 34.01 & $-2.05$ & $-0.27$ & $-3.31$\\%34.5036, 41.640358535673045, 34.00595987367331
Flat \pcdm & H$\beta$ mono & $\cdots$ & $\cdots$ & 0.995 & $\cdots$ & 7.726 & $\cdots$ & 0.433 & 1.373 & 0.197 & $-10.92$ & $-2.92$ & 9.31 & $-2.23$ & 1.99 & 5.05 & $-1.23$\\%-2.91848, 9.30650322139323, -2.2299082592406485
 & $H(z)$ + BAO + H$\beta$ mono & 0.0353 & 0.0804 & 0.265 & $\cdots$ & 1.469 & 66.21 & 0.422 & 1.342 & 0.199 & 19.45 & 33.45 & 56.57 & 33.05 & $-3.19$ & 0.12 & $-4.49$\\%33.45098, 56.57411435641353, 33.046576942002105
[6pt]
 & $H(z)$ + BAO & 0.0338 & 0.0878 & 0.273 & $-0.077$ & 1.441 & 66.86 & $\cdots$ & $\cdots$ & $\cdots$ & 25.92 & 35.92 & 44.84 & 35.12 & $-0.64$ & 2.93 & $-2.20$\\%35.9212, 44.842148169591304, 35.12310308752016 D
Nonflat \pcdm & H$\beta$ mono & $\cdots$ & $\cdots$ & 0.967 & $-0.951$ & 9.931 & $\cdots$ & 0.443 & 1.387 & 0.195 & $-12.41$ & $-2.41$ & 12.87 & $-1.24$ & 2.50 & 8.61 & $-0.24$\\%-2.40752, 12.873709026741539, -1.241612972212252 D
 & $H(z)$ + BAO + H$\beta$ mono & 0.0367 & 0.0781 & 0.261 & $-0.096$ & 1.860 & 66.57 & 0.423 & 1.347 & 0.201 & 18.90 & 34.90 & 61.33 & 34.04 & $-1.73$ & 4.87 & $-3.50$\\%34.90448, 61.330919264472605, 34.037102225406834 D
\bottomrule\bottomrule
\end{tabular}
%}
\begin{tablenotes}
\item [a] \wx\ corresponds to flat/nonflat XCDM and $\alpha$ corresponds to flat/nonflat \pcdm.
\item [b] \hunit.
% \item [c] $\Omega_b=0.05$ and $H_0=70$ \hunit.
\end{tablenotes}
\end{threeparttable}%
}
\end{table*}
\end{turnpage}

\begin{turnpage}
\begin{table*}
\centering
% \footnotesize
% \resizebox*{2.7\columnwidth}{1.2\columnwidth}{%
\resizebox{2.7\columnwidth}{!}{%
\begin{threeparttable}
\caption{One-dimensional marginalized posterior mean values and uncertainties ($\pm 1\sigma$ error bars or $2\sigma$ limits) of the parameters for all models from various combinations of data, where H$\beta$ mono stands for H$\beta$ time delays in combination with the monochromatic luminosity at 5100 \AA.}\label{tab:1d_BFP}
\begin{tabular}{lcccccccccc}
\toprule\toprule
Model & dataset & $\Omega_{b}h^2$ & $\Omega_{c}h^2$ & $\Omega_{m0}$ & $\Omega_{k0}$ & $w_{\mathrm{X}}$/$\alpha$\tnote{a} & $H_0$\tnote{b} & $\gamma$ & $\beta$ & $\sigma_{\mathrm{int}}$\\
\midrule
 & $H(z)$ + BAO & $0.0260\pm0.0040$ & $0.1213^{+0.0091}_{-0.0103}$ & $0.298^{+0.015}_{-0.018}$ & $\cdots$ & $\cdots$ & $70.51\pm2.72$ & $\cdots$ & $\cdots$ & $\cdots$ \\
Flat \lcdm & H$\beta$ mono & $\cdots$ & $\cdots$ & $>0.319$ & $\cdots$ & $\cdots$ & $\cdots$ & $0.428\pm0.025$ & $1.368\pm0.022$ & $0.202^{+0.015}_{-0.017}$ \\
 & $H(z)$ + BAO + H$\beta$ mono & $0.0258^{+0.0036}_{-0.0040}$ & $0.1216^{+0.0088}_{-0.0099}$ & $0.299^{+0.015}_{-0.017}$ & $\cdots$ & $\cdots$ & $70.40\pm2.62$ & $0.418^{+0.023}_{-0.024}$ & $1.356\pm0.024$ & $0.205^{+0.014}_{-0.017}$ \\
[6pt]
 & $H(z)$ + BAO & $0.0275^{+0.0047}_{-0.0053}$ & $0.1132\pm0.0183$ & $0.289\pm0.023$ & $0.047^{+0.083}_{-0.091}$ & $\cdots$ & $69.81\pm2.87$ & $\cdots$ & $\cdots$ & $\cdots$ \\
Nonflat \lcdm & H$\beta$ mono & $\cdots$ & $\cdots$ & $>0.196$ & $>-0.362$ & $\cdots$ & $\cdots$ & $0.437\pm0.026$ & $1.383\pm0.024$ & $0.201^{+0.014}_{-0.017}$ \\
 & $H(z)$ + BAO + H$\beta$ mono & $0.0275^{+0.0046}_{-0.0052}$ & $0.1128\pm0.0179$ & $0.289\pm0.023$ & $0.055^{+0.082}_{-0.090}$ & $\cdots$ & $69.62\pm2.80$ & $0.418\pm0.023$ & $1.353\pm0.024$ & $0.205^{+0.014}_{-0.017}$ \\
[6pt]
 & $H(z)$ + BAO & $0.0308^{+0.0053}_{-0.0046}$ & $0.0980^{+0.0182}_{-0.0161}$ & $0.286\pm0.019$ & $\cdots$ & $-0.778^{+0.132}_{-0.104}$ & $67.20^{+3.05}_{-3.06}$ & $\cdots$ & $\cdots$ & $\cdots$ \\
Flat XCDM & H$\beta$ mono & $\cdots$ & $\cdots$ & $>0.220$ & $\cdots$ & $-1.758^{+1.917}_{-0.855}$ & $\cdots$ & $0.428\pm0.025$ & $1.368\pm0.022$ & $0.202^{+0.015}_{-0.017}$ \\
 & $H(z)$ + BAO + H$\beta$ mono & $0.0315^{+0.0056}_{-0.0043}$ & $0.0944^{+0.0183}_{-0.0163}$ & $0.284\pm0.020$ & $\cdots$ & $-0.747^{+0.127}_{-0.095}$ & $66.66^{+2.94}_{-2.95}$ & $0.421\pm0.024$ & $1.341\pm0.025$ & $0.204^{+0.014}_{-0.017}$ \\
[6pt]
 & $H(z)$ + BAO & $0.0305^{+0.0055}_{-0.0047}$ & $0.1011\pm0.0196$ & $0.292\pm0.024$ & $-0.059\pm0.106$ & $-0.746^{+0.135}_{-0.090}$ & $67.17^{+2.96}_{-2.97}$ & $\cdots$ & $\cdots$ & $\cdots$ \\
Nonflat XCDM & H$\beta$ mono & $\cdots$ & $\cdots$ & $>0.543$\tnote{c} & $>0.902$\tnote{c} & $<0.011$ & $\cdots$ & $0.448\pm0.028$ & $1.406^{+0.031}_{-0.035}$ & $0.200^{+0.014}_{-0.017}$ \\
 & $H(z)$ + BAO + H$\beta$ mono & $0.0311^{+0.0059}_{-0.0044}$ & $0.0979^{+0.0197}_{-0.0198}$ & $0.290\pm0.024$ & $-0.067\pm0.108$ & $-0.717^{+0.132}_{-0.082}$ & $66.70^{+2.90}_{-2.92}$ & $0.421\pm0.024$ & $1.341\pm0.025$ & $0.204^{+0.014}_{-0.017}$ \\
[6pt]
 & $H(z)$ + BAO & $0.0327^{+0.0060}_{-0.0031}$ & $0.0866^{+0.0192}_{-0.0176}$ & $0.272\pm0.022$ & $\cdots$ & $1.261^{+0.494}_{-0.810}$ & $66.23^{+2.87}_{-2.86}$ & $\cdots$ & $\cdots$ & $\cdots$ \\%$1.261^{+1.276}_{-1.206}$ 2$\sigma$
Flat \pcdm & H$\beta$ mono & $\cdots$ & $\cdots$ & $>0.191$ & $\cdots$ & $>3.868$\tnote{c} & $\cdots$ & $0.430\pm0.025$ & $1.371\pm0.021$ & $0.202^{+0.014}_{-0.017}$ \\
 & $H(z)$ + BAO + H$\beta$ mono & $0.0332^{+0.0066}_{-0.0020}$ & $0.0841\pm0.0187$ & $0.270\pm0.023$ & $\cdots$ & $1.382^{+0.519}_{-0.843}$ & $65.93\pm2.84$ & $0.421\pm0.024$ & $1.337\pm0.025$ & $0.204^{+0.014}_{-0.017}$ \\
[6pt]
 & $H(z)$ + BAO & $0.0324^{+0.0062}_{-0.0031}$ & $0.0900\pm0.0200$ & $0.277\pm0.025$ & $-0.072^{+0.093}_{-0.107}$ & $1.435^{+0.579}_{-0.788}$ & $66.50\pm2.88$ & $\cdots$ & $\cdots$ & $\cdots$ \\%$1.435^{+1.319}_{-1.290}$
Nonflat \pcdm & H$\beta$ mono & $\cdots$ & $\cdots$ & $>0.544$\tnote{c} & $-0.166^{+0.353}_{-0.440}$ & $>4.078$\tnote{c} & $\cdots$ & $0.430\pm0.024$ & $1.372\pm0.021$ & $0.202^{+0.014}_{-0.017}$ \\
 & $H(z)$ + BAO + H$\beta$ mono & $0.0329^{+0.0068}_{-0.0021}$ & $0.0876\pm0.0203$ & $0.275\pm0.025$ & $-0.073^{+0.091}_{-0.108}$ & $1.552^{+0.607}_{-0.818}$ & $66.22\pm2.85$ & $0.421\pm0.023$ & $1.339\pm0.025$ & $0.204^{+0.014}_{-0.017}$ \\
\bottomrule\bottomrule
\end{tabular}
\begin{tablenotes}
\item [a] \wx\ corresponds to flat/nonflat XCDM and $\alpha$ corresponds to flat/nonflat \pcdm.
\item [b] \hunit. For the RM AGN alone cases, $\Omega_b=0.05$ and $H_0=70$ \hunit.
\item [c] This is the 1$\sigma$ limit. The 2$\sigma$ limit is set by the prior and not shown here.
\end{tablenotes}
\end{threeparttable}%
}
\end{table*}
\end{turnpage}

Table~\ref{tab:diff1} highlights the maximum variations in the $R-L$ relation and intrinsic scatter parameters across the tested cosmological models, suggesting that the H$\beta$ mono RM AGN dataset can be standardized using the $R-L$ relation. Note that excluding the nonflat \lcdm\ and XCDM models, which are observationally inconsistent unless \ok\ is exceedingly small \citep{deCruzPerez:2024shj}, strengthens this conclusion. \footnote{Using Planck CMB anisotropy data in conjunction with BAO, $H(z)$, supernova Type Ia, and other non-CMB data, in the nonflat $\Lambda$CDM model one measures $\Omega_{k0} = 0.0009 \pm 0.0017$ or $\Omega_{k0} = 0.0008 \pm 0.0017$, depending on the assumed primordial inhomogeneity power spectrum, \citep{deCruzPerez:2024shj}. In the nonflat XCDM case the constraints from Planck CMB data and those from the combined non-CMB dataset are inconsistent at $> 3\sigma$ and, therefore, nonflat XCDM is observationally ruled out, \citep{deCruzPerez:2024shj}. Extending the nonflat XCDM model by also allowing the gravitational lensing consistency parameter $A_L$ to vary and be determined by these data brings the CMB and non-CMB data constraints into $< 3\sigma$ inconsistency and when jointly analyzed these give $\Omega_{k0} = 0.0015 \pm 0.0019$ for both primordial inhomogeneity power spectra, \citep{deCruzPerez:2024shj}. For our purposes the relevant point is that in the nonflat $\Lambda$CDM and XCDM cases these data require that the models be very close to, if not exactly, spatially flat and, therefore, essentially indistinguishable from the flat $\Lambda$CDM and XCDM cases.}

\begin{table}
\centering
% \resizebox{2\columnwidth}{!}{%
\setlength\tabcolsep{17pt}
\begin{threeparttable}
\caption{The largest differences between results for considered cosmological models from H$\beta$ mono with $1\sigma$ being the quadrature sum of the two corresponding $1\sigma$ error bars.}\label{tab:diff1}
% various combinations of data 
%\setlength{\tabcolsep}{0.5mm}{
\begin{tabular}{lccc}
\toprule\toprule
 Data set & $\Delta\gamma$ & $\Delta\beta$ & $\Delta\sigma_{\mathrm{int}}$ \\
\midrule
H$\beta$ mono\tnote{a} & $0.06\sigma$ & $0.13\sigma$ & $0$\\
H$\beta$ mono\tnote{b} & $0.53\sigma$ & $0.92\sigma$ & $0.09\sigma$\\
\bottomrule\bottomrule
\end{tabular}
%}
\begin{tablenotes}[flushleft]
\item [a] Flat \lcdm\ and XCDM, and flat and nonflat \pcdm\ models.
\item [b] All six flat and nonflat \lcdm, XCDM, and \pcdm\ models.
\end{tablenotes}
\end{threeparttable}%
% }
\end{table}

For H$\beta$ mono data alone, the slope ($\gamma$) varies from $0.428\pm0.025$ (flat \lcdm\ and XCDM) to $0.448\pm0.028$ (nonflat XCDM). The intercept ($\beta$) ranges from $1.368\pm0.022$ (flat \lcdm\ and XCDM) to $1.406^{+0.031}_{-0.035}$ (nonflat XCDM). The intrinsic scatter ($\sigma_{\rm int}$) spans from $0.200^{+0.014}_{-0.017}$ (nonflat XCDM) to $0.202^{+0.015}_{-0.017}$ (flat \lcdm\ and XCDM). The intrinsic scatter remains stable across models. 

Comparing the 157 sources sample results here with the earlier 41 sources of H$\beta$ mono data from \citet{Caoetal2025}, we see that the median slope $\gamma$ error bars here, 0.025, are significantly smaller than the Ref.~\citep{Caoetal2025} median value, 0.059, with the median intercept $\beta$ error bar here, 0.022, also significantly smaller than the corresponding Ref.~\citep{Caoetal2025} value, 0.057, and the median intrinsic scatter $\sigma_{\rm int}$ here, 0.202, is smaller than that of Ref.~\citep{Caoetal2025}, 0.270. Unlike the 41 source case studied in Ref.~\citep{Caoetal2025} where we found a steeper slope than the $\gamma = 0.5$ expected in a simple photoionization model of BLR clouds \citep{1972ApJ...171..213D,2019FrASS...6...75P,2021bhns.confE...1K}, steeper by $1.4\sigma$, here we find a shallower slope, $2.8-2.9\sigma$ smaller than $\gamma = 0.5$ for flat \lcdm, flat XCDM, and flat and nonflat \pcdm, and $2.4\sigma$ and $1.9\sigma$ smaller for nonflat \lcdm\ and nonflat XCDM.

Figure~\ref{fig2} shows that the H$\beta$ mono dataset of 157 sources favors a currently decelerating cosmological expansion. The likelihood peaks in the decelerating region of parameter space, although accelerating expansion remains consistent within $2\sigma$, as the $2\sigma$ contours still overlap part of the accelerating region.

The H$\beta$ mono data alone constrain key cosmological parameters as follows. Among the six cosmological models, the $1\sigma$ lower limit on \om\ reaches as low as $> 0.543$ in nonflat XCDM (with the $2\sigma$ limit being $> 0.0513$, determined by the prior), while the highest $2\sigma$ lower bound is $> 0.319$ in flat \lcdm; we quote these two extremes to illustrate the full range of \om\ lower limits across models. The \ok\ constraints are $>-0.362$ ($2\sigma$), $>0.902$ ($1\sigma$), and $-0.166^{+0.353}_{-0.440}$ for nonflat \lcdm, XCDM, and \pcdm, respectively. For the XCDM parametrizations, the \wx\ constraints are $-1.758^{+1.917}_{-0.855}$ and $<0.011$ ($2\sigma$) for the flat and nonflat cases, respectively; whereas for \pcdm\ models, the $\alpha$ constraints are $>3.868$ ($1\sigma$) and $>4.078$ ($1\sigma$) for the flat and nonflat cases, respectively.

Given that H$\beta$ mono data can be standardized through the $R-L$ relation and its derived 1D marginalized cosmological parameters are consistent with those from $H(z)$ + BAO data, a joint analysis of $H(z)$ + BAO + H$\beta$ mono data might be justified. However, a tension in the $\Om-\Ok$ plane for nonflat \lcdm\ and nonflat XCDM models, where the $H(z)$ + BAO data and the H$\beta$ mono data $2\sigma$ confidence regions do not overlap, indicates a potential incompatibility between these data in these two models and argues against a joint analysis of these two datasets under these two models. We note, however, that in both of these models a combined analysis of CMB and non-CMB data, which carry significantly more statistical weight than data used in this manuscript, indicates that $\Omega_{k0}$ must be very small, making the two nonflat cases almost indistinguishable from their flat counterparts \citep{deCruzPerez:2024shj}.

A joint analysis of $H(z)$ + BAO + H$\beta$ mono data results in only minor shifts in the cosmological constraints compared to those from just $H(z)$ + BAO data. The most significant changes are a $0.072\sigma$ shift in \om\ for flat XCDM, a $0.19\sigma$ shift in \wx\ for flat XCDM, a $0.12\sigma$ shift in $\alpha$ for flat \pcdm, and a $0.13\sigma$ shift in $H_0$ for flat XCDM.

\begin{table}
\centering
% \resizebox{2\columnwidth}{!}{%
\setlength\tabcolsep{13pt}
\begin{threeparttable}
\caption{The differences between H$\beta$ mono and $H(z)$ + BAO + H$\beta$ mono results for a given cosmological model with $1\sigma$ being the quadrature sum of the two corresponding $1\sigma$ error bars.}\label{tab:diff2}
\begin{tabular}{lccc}
\toprule\toprule
 Model & $\Delta\gamma$ & $\Delta\beta$ & $\Delta\sigma_{\mathrm{int}}$ \\
\midrule
Flat \lcdm & $0.29\sigma$ & $0.37\sigma$ & $0.13\sigma$ \\
Nonflat \lcdm & $0.55\sigma$ & $0.88\sigma$ & $0.18\sigma$ \\
% Flat XCDM & $0.20\sigma$ & $0.81\sigma$ & $0.088\sigma$ \\
Flat XCDM & $0.20\sigma$ & $0.81\sigma$ & $0.09\sigma$ \\
Nonflat XCDM & $0.73\sigma$ & $1.51\sigma$ & $0.18\sigma$ \\
% Flat \pcdm & $0.26\sigma$ & $1.04\sigma$ & $0.091\sigma$ \\
Flat \pcdm & $0.26\sigma$ & $1.04\sigma$ & $0.09\sigma$ \\
% Nonflat \pcdm & $0.27\sigma$ & $1.01\sigma$ & $0.091\sigma$ \\
Nonflat \pcdm & $0.27\sigma$ & $1.01\sigma$ & $0.09\sigma$ \\
\bottomrule\bottomrule
\end{tabular}
%}
% \begin{tablenotes}[flushleft]
% \item [a] Only for flat/Nonflat \lcdm.

% \end{tablenotes}
\end{threeparttable}%
% }
\end{table}

As is listed in Table~\ref{tab:diff2}, combining H$\beta$ mono with $H(z)$ + BAO data highlights notable differences in the $R-L$ relation and intrinsic scatter parameters. The largest differences in the slope $\gamma$ and intercept $\beta$ are $0.73\sigma$ and $1.51\sigma$, respectively, both occurring under the nonflat XCDM parametrization. The largest difference in the intrinsic scatter $\sigma_{\rm int}$ is $0.18\sigma$, observed under nonflat \lcdm\ and XCDM.

Based on the more reliable DIC, H$\beta$ mono data alone favors nonflat XCDM the most, showing weak or positive evidence against other models and parametrizations. In contrast, the joint $H(z)$ + BAO + H$\beta$ mono dataset favors flat \pcdm\ the most, showing also weak or positive evidence against other models and parametrizations. However, in all the cases $\Delta$DIC values do not strongly distinguish between any two of the models.

\begin{figure*}[htbp]
\centering
 \subfloat[Flat \lcdm]{%
    \includegraphics[width=0.45\textwidth,height=0.35\textwidth]{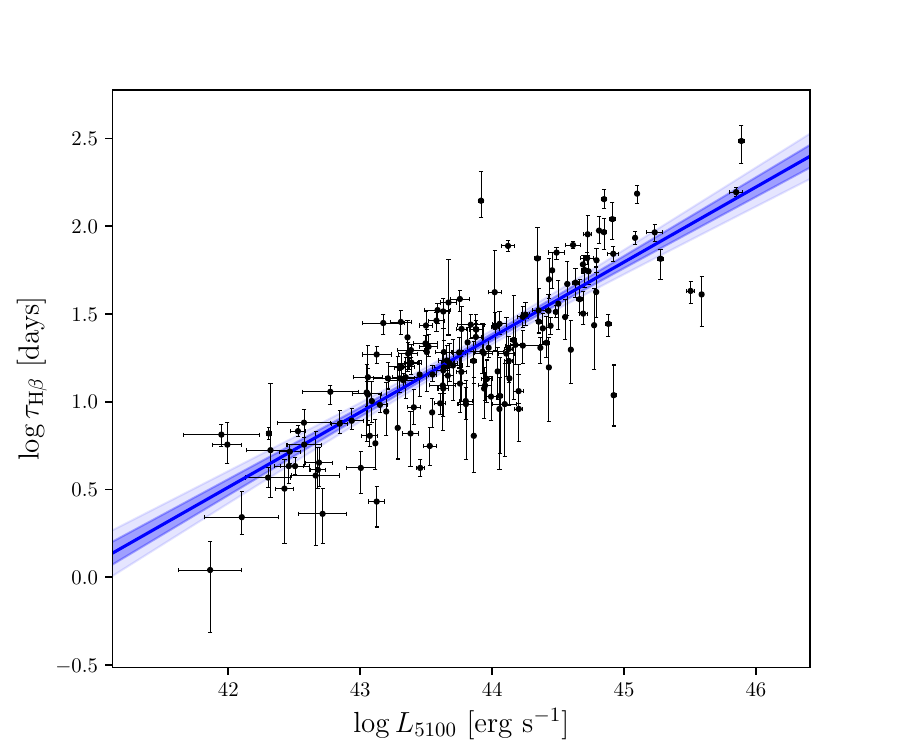}}
 \subfloat[Nonflat \lcdm]{%
    \includegraphics[width=0.45\textwidth,height=0.35\textwidth]{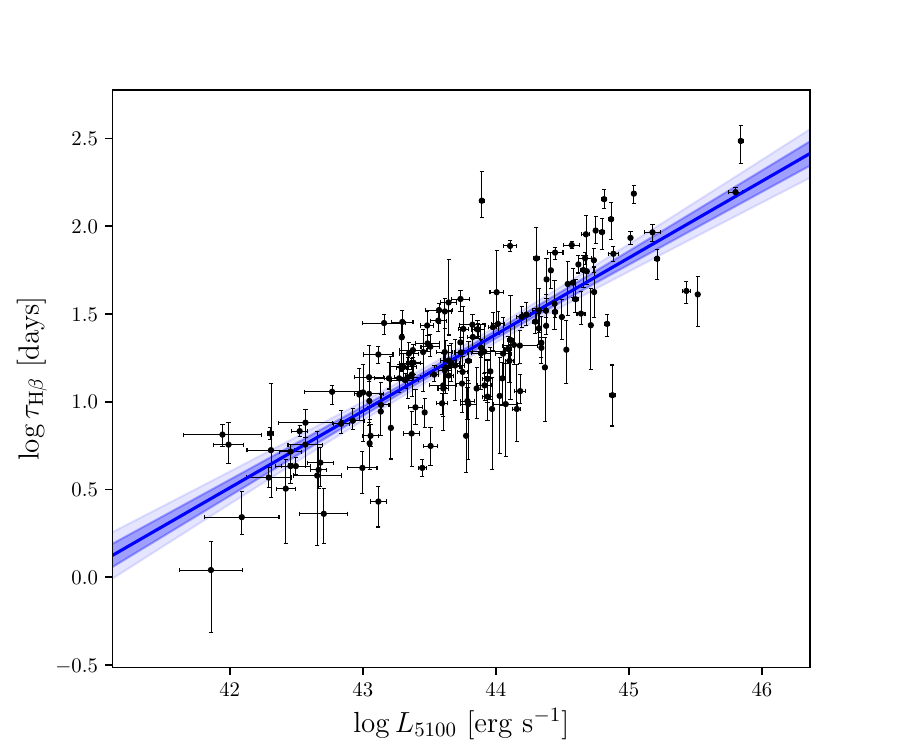}}\\
 \subfloat[Flat XCDM]{%
    \includegraphics[width=0.45\textwidth,height=0.35\textwidth]{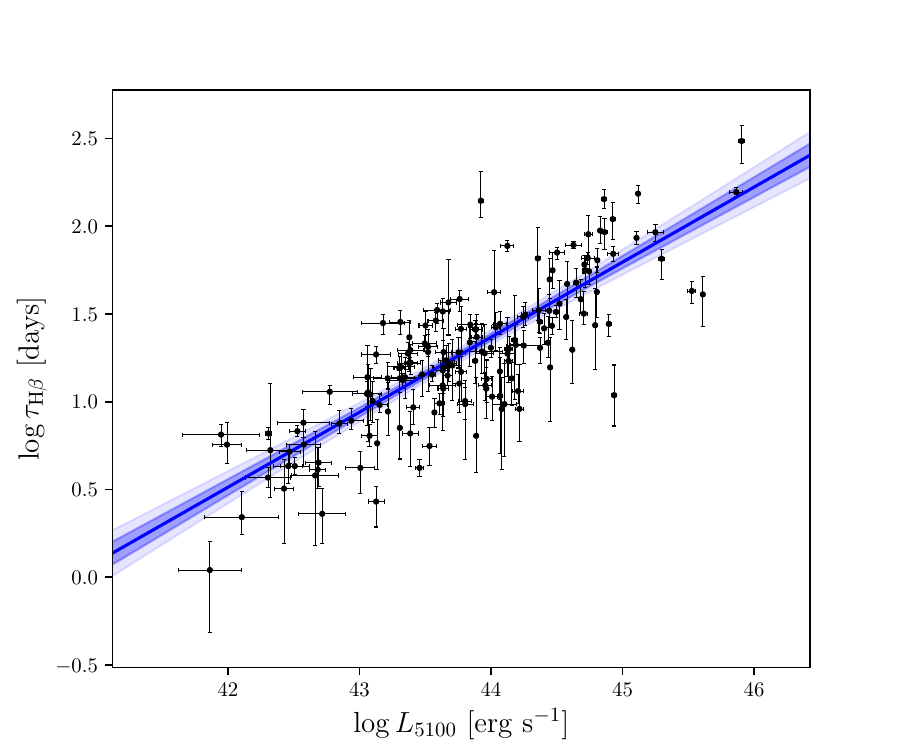}}
 \subfloat[Nonflat XCDM]{%
    \includegraphics[width=0.45\textwidth,height=0.35\textwidth]{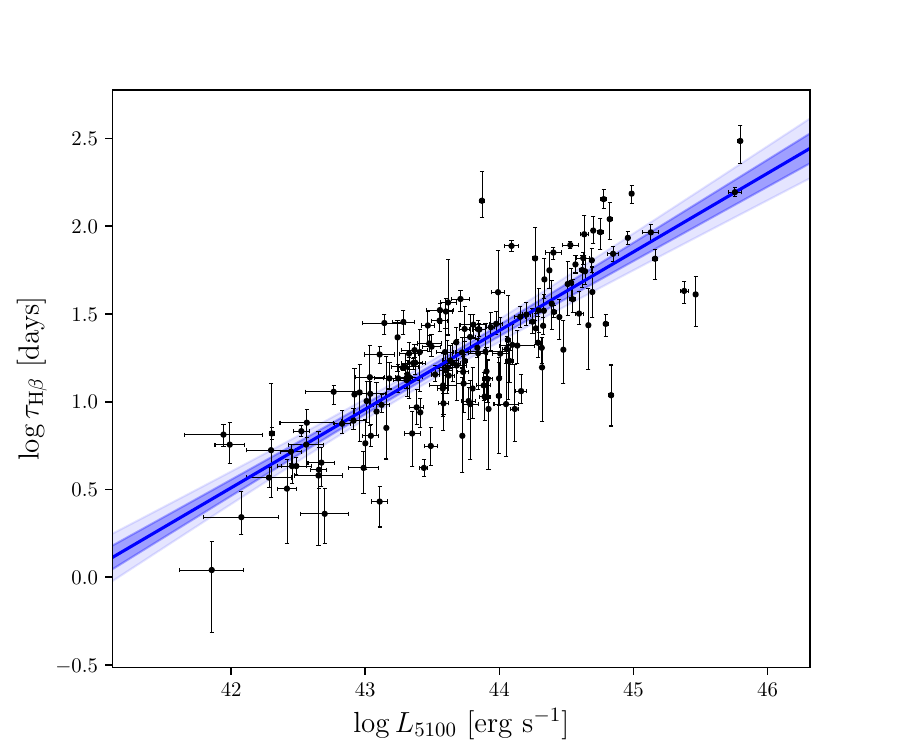}}\\
 \subfloat[Flat \pcdm]{%
    \includegraphics[width=0.45\textwidth,height=0.35\textwidth]{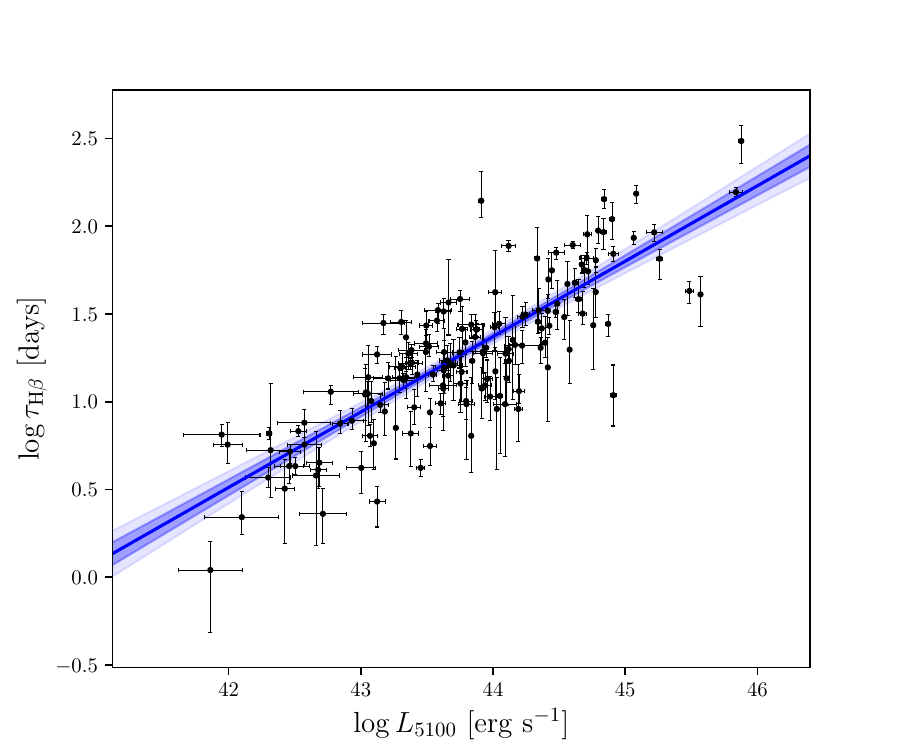}}
 \subfloat[Nonflat \pcdm]{%
    \includegraphics[width=0.45\textwidth,height=0.35\textwidth]{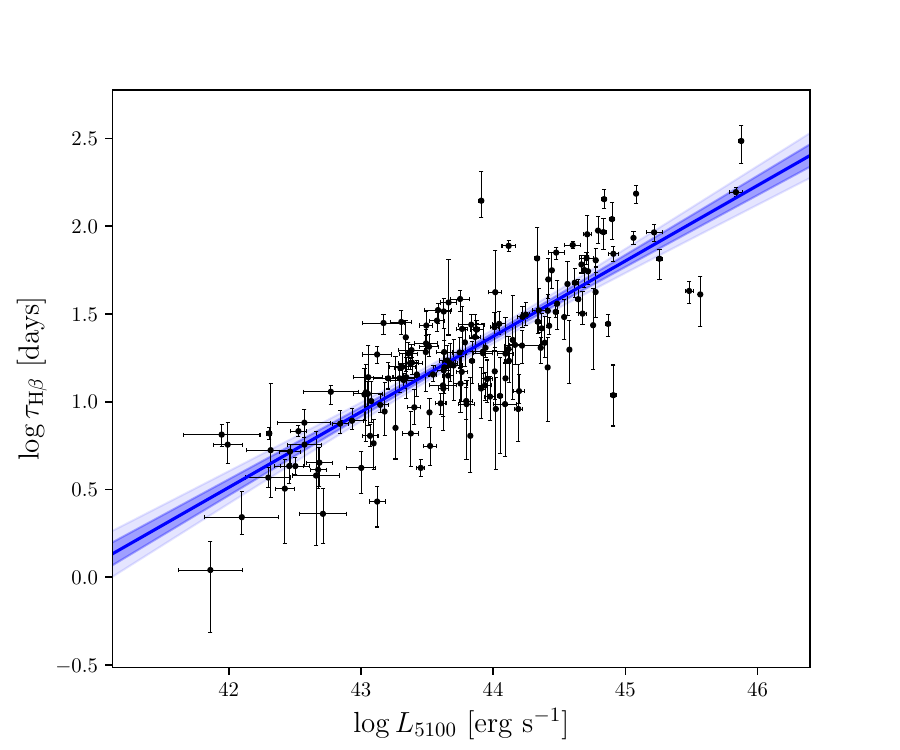}}\\
\caption{$R-L$ relations of H$\beta$ mono data for six different models, with \lcdm, XCDM, and \pcdm\ in the top, middle, and bottom rows, and flat (nonflat) models in the left (right) column. Here, $D_L$ values for the data points are computed using the corresponding posterior mean values listed in Table \ref{tab:1d_BFP}. The $R-L$ relations themselves are derived from Monte Carlo simulations, using the posterior mean values from Table \ref{tab:1d_BFP} and covariance matrices of the intercept and slope parameters. Errors of $\log \tau$ are from error propagation of $\tau$.}
\label{R-L1}
\vspace{-50pt}
\end{figure*}

Figure \ref{R-L1} shows the H$\beta$ mono $R-L$ relations for six different cosmological models. The black data points represent luminosities computed using Eq.~\eqref{eq:L5100} with posterior mean values of the cosmological parameters provided in Table \ref{tab:1d_BFP}. The $R-L$ relations are derived from Monte Carlo simulations, incorporating the posterior mean values and covariance matrices of the intercept and slope parameters, also presented in Table \ref{tab:1d_BFP}. The error bars for $\log \tau_{\text{H}\beta}$ are obtained through error propagation of $\tau$. Notably, the observed data points generally show good agreement with the predicted $R-L$ relations.

\subsection{BLR radius$-$luminosity relation}
\label{sec:r_l}

\begin{figure}
\centering
\includegraphics[scale=0.7]{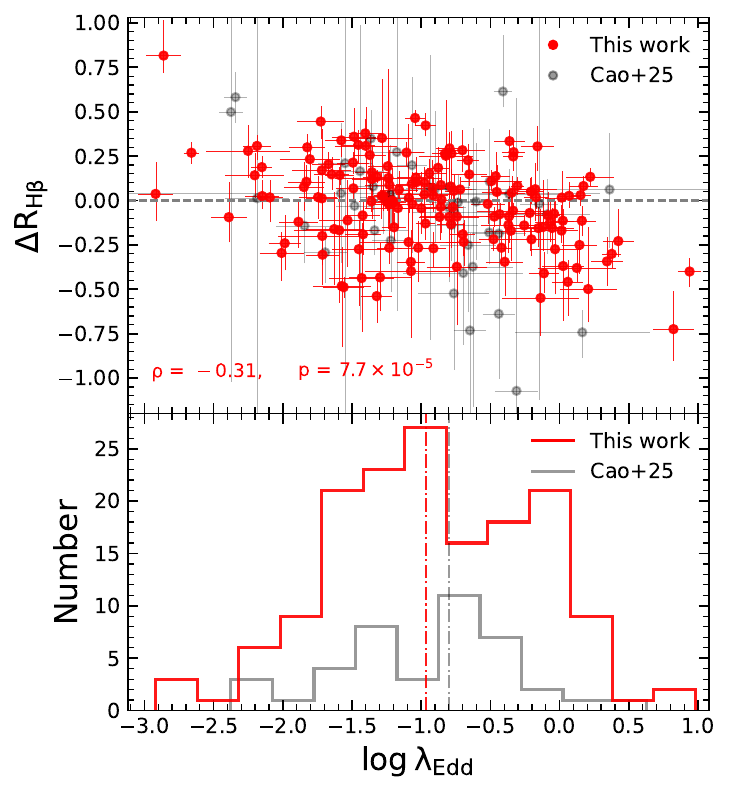}
%\resizebox{6cm}{5cm}{\includegraphics{grt_curve.pdf}}
\caption{Top: Deviations in the BLR radii based on H$\beta$ lines from the best-fitting $R-L$ relation ($\Delta R_\mathrm{H\beta}$) plotted against the Eddington ratio. Red circular points represent measurements from this work, while black points correspond to those from \citet{Caoetal2025}. The Spearman's rank correlation coefficients and their associated probabilities are indicated. The gray dashed horizontal line marks $\Delta R_\mathrm{H\beta} = 0$.
Bottom: Histogram distribution of the Eddington ratio for the H$\beta$ mono sample used in this work (red) and the sample from \citet{Caoetal2025} (black) using the same uniform bin size for both samples. The median values of each distribution are indicated by dot-dashed vertical lines. The current sample significantly increases the number of data points and covers a broader range in the Eddington ratio, including a larger number of high-accreting AGNs compared to the previous sample.}
\label{fig:scat}
\end{figure}

With an increasing number of H$\beta$ RM measurements in AGNs, the canonical $R-L$ relation, proposed by \citet{2013ApJ...767..149B} with a slope of approximately 0.5 as predicted in a simple photoionization model, now reveals greater complexity and a larger scatter. Previous analyses of RM data \citep{2016ApJ...825..126D,2017ApJ...851...21G,MartinezAldama2019} revealed that this scatter can be linked to the Eddington ratio ($\lambda_{\rm Edd}$) of the sources or the related relative accretion rate and quantities that generally serve as proxies of the relative accretion rate, such as the relative \Feii\ strength \citep{2019ApJ...886...42D}. Other contributing factors could be sample selection due to a monitoring cadence \citep{2017ApJ...851...21G} as well as the general shape of the UV/optical spectral energy distribution and the associated relative fraction of ionizing photons \citep{2020ApJ...899...73F}.  

Using a revised, high-quality Best sample of 157 H$\beta$ RM measurements, \citet{2024ApJ...962...67W} and \citet{2024ApJS..275...13W} confirm that super-Eddington AGNs ($\lambda_{\rm Edd} > 1$) exhibit significantly shortened H$\beta$ radii compared to expectations, resulting in deviations from the canonical $R-L$ relation. This finding is consistent with earlier results reported by \citet{2015ApJ...806...22D, 2018ApJ...856....6D}, and a similar trend is also observed in the torus radius$-$luminosity relation in AGNs \citep{2024ApJ...968...59M}.

Therefore, we investigate the $R-L$ relations for the Best sample under six different cosmological models using Eq.~\eqref{eq:R-L}, as shown in Fig.\ \ref{R-L1}. We find consistent slope values within uncertainties, ranging from $\gamma$ = 0.428 to 0.430 with an uncertainty of approximately 0.025, across most cosmological models. The exceptions are the nonflat $\Lambda$CDM and XCDM models, which show slightly larger slopes of $0.437\pm0.026$ and $0.448\pm0.028$, respectively. Similarly, the intercept values remain consistent among different cosmologies, with $\beta$ ranging between $1.368\pm0.022$ and $1.372\pm0.021$, except again for the nonflat $\Lambda$CDM ($1.383\pm0.024$) and XCDM ($1.406^{+0.031}_{-0.035}$) models. Aside from these two exceptions, which as noted above are inconsistent with a large compilation of CMB and non-CMB data unless they are close to or spatially flat, the obtained slopes and intercepts agree well with the values reported by \citet{2024ApJS..275...13W}, who found $\gamma = 0.42\pm0.02$ and $\beta=1.34\pm0.02$. 

Additionally, as noted above, we observe a significantly shallower slope in the $R-L$ relation in this work compared to both the value reported by \citet{Caoetal2025} and the theoretically expected slope of 0.5 from photoionization model. For instance, \citet{Caoetal2025} found a slope of $\gamma = 0.584 \pm 0.059$, which is notably steeper than our measured value of $0.428 \pm 0.025$ for the flat $\Lambda$CDM model. 

To explore the origin of this discrepancy, we examine the scatter in the $R-L$ relation, defined as the deviation of the measured BLR radii from the best-fit relation, $\Delta R_{\rm{H\beta}} = {\rm log}R - {\rm log}R_{\text{best-fit}}$ as a function of accretion rate defined by the Eddington ratios, $\lambda_{\text{Edd}}$, of the sources. To compute $\lambda_{\text{Edd}}$ we first obtain the velocity dispersion $\Delta V$ of the H$\beta$ line either from the line dispersion ($\sigma_{\text{line}}$) or from the full width at half maximum (FWHM) when $\sigma_{\text{line}}$ is not available from  \citet{2024ApJS..275...13W} and \citet{2024ApJ...962...67W}. Using this velocity estimate, we compute the black hole mass ($M_{\text{BH}}$) via the virial method, adopting a virial factor $f_{\text{BLR}} = 4.47$ when using $\sigma_{\text{line}}$ or $f_{\text{BLR}} = 1.12$ for FWHM \citep{2015ApJ...801...38W}, from $M_{\text{BH}}$ = $f_{\text{BLR}} (c\tau_{\rm{H}\beta} \Delta V^2/{G})$. Next, we compute the bolometric luminosity ($L_{\text{\text{bol}}}$) by applying a bolometric correction factor of 9.26 to $L_{5100}$, \citep{2006ApJS..166..470R}, for the flat $\Lambda$CDM model. Subsequently, the Eddington luminosity is computed using $L_{\text{Edd}}$ = $1.26 \times 10^{38} M_{\text{BH}}/M_{\odot}$ and the Eddington ratio is obtained from $\lambda_{\text{Edd}} = L_{\text{bol}}/L_{\text{Edd}}$. 

We plot $\Delta R_{\rm{H\beta}}$ against  $\lambda_{\text{Edd}}$ in the top panel of Fig.~\ref{fig:scat}. This plot clearly reveals a trend of decreasing BLR radius with increasing $\lambda_{\text{Edd}}$ (see also \citet{Caoetal2025} and \citet{2024ApJS..275...13W} for more details). Furthermore, the Eddington ratio distributions shown in the bottom panel of Fig.~\ref{fig:scat} indicate that our current H$\beta$ mono sample includes a greater number of high-accreting AGNs. These findings suggest that the shallower slope observed in our analysis is primarily driven by the inclusion of a significantly larger sample with a broader range of Eddington ratios, particularly enriched in sources with higher accretion rates.

\begin{figure*}
    \centering
    \includegraphics[width=0.32\textwidth]{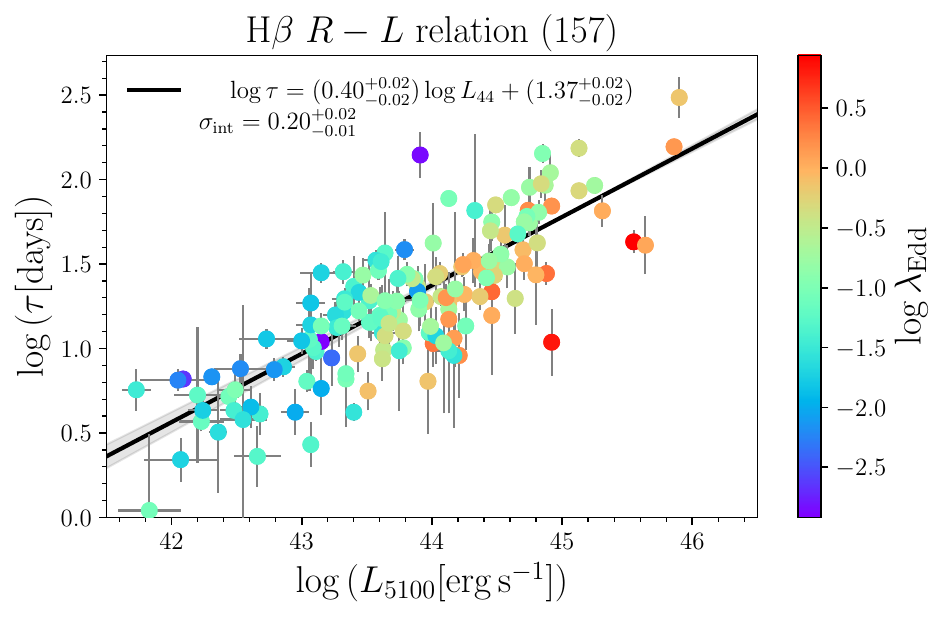}
    \includegraphics[width=0.32\textwidth]{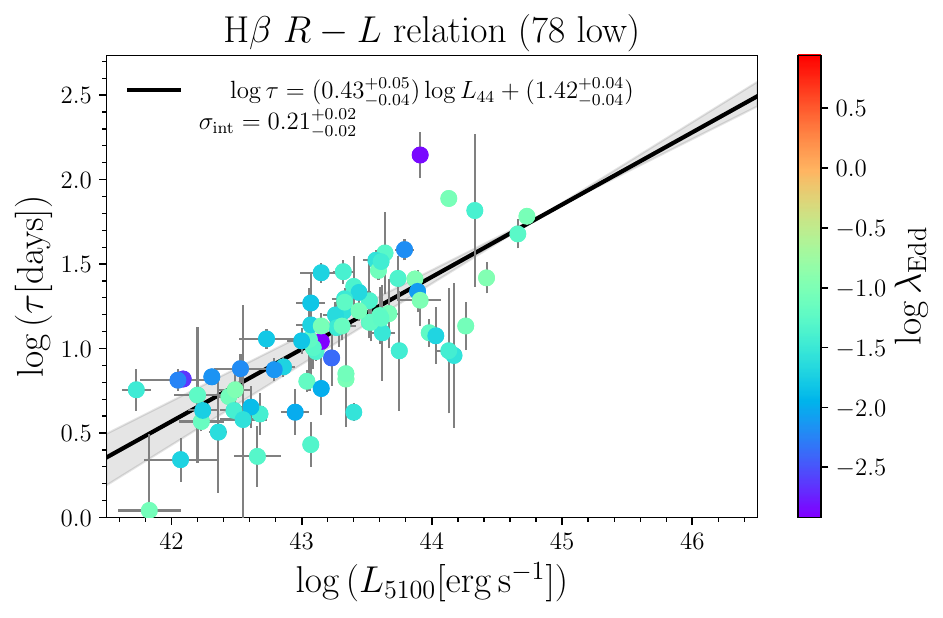}
    \includegraphics[width=0.32\textwidth]{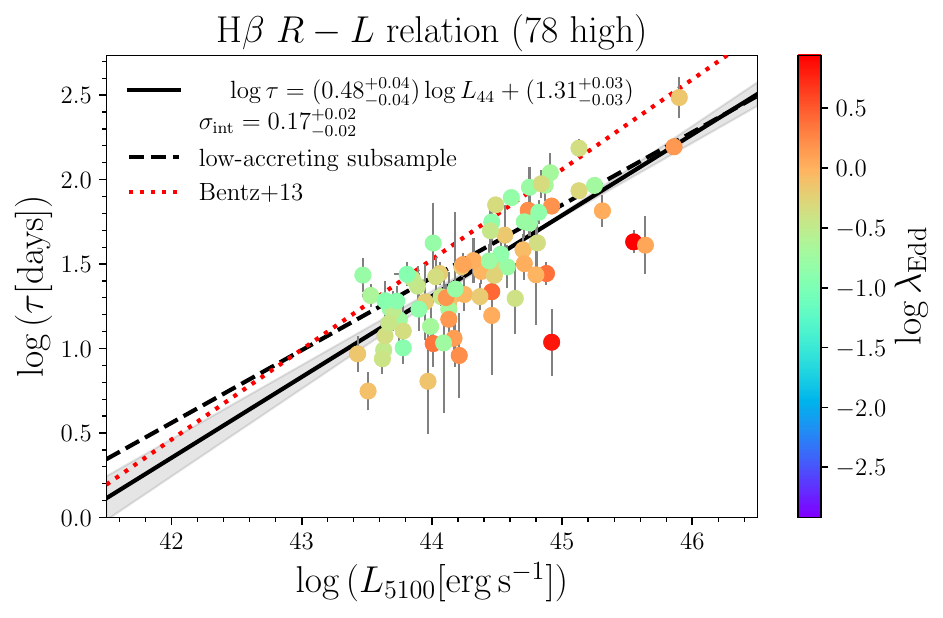}
    \caption{H$\beta$ $R-L$ relations in the fixed flat $\Lambda$CDM model (with $H_0=72\,{\rm km\,s^{-1}\,Mpc^{-1}}$ and $\Omega_{\rm m0}=0.3$). We show the $R-L$ relation for whole sample of 157 sources (left panel). In addition, we depict the $R-L$ relations for the low-accreting subsample of 78 sources (middle panel) as well as for the equal-sized subsample of high-accreting sources (right panel). Each data point is color-coded by the logarithm of the Eddington ratio, $\lambda_{\rm Edd}$ with the color axis to the right of each panel. The best-fit $R-L$ relations are indicated by black solid lines and the parameters, including the intrinsic scatter, are included in the legend. The gray-shaded areas around the best-fit relations correspond to 1$\sigma$ uncertainties of the best fit. In the right panel, we also depict the best-fit $R-L$ relation for the low-accreting subsample from the middle panel for comparison (dashed line) as well as the Bentz et al. (2013) H$\beta$ $R-L$ relation \citep{2013ApJ...767..149B} that is very close to the values expected from the photoionization model. We keep the axes ranges the same for all three $R-L$ relations for ease of comparison.}
    \label{fig_RLrel_lowhigh}
\end{figure*}

Another way to approach the effect of the relative accretion rate is to look at the $R-L$ relations separately for the low- and high-accreting sources. We perform a test by dividing the sample into two halves with respect to the median value of the logarithm of the Eddington ratio, which is $\log{\lambda_{\rm Edd}}=-0.9683$. This Eddington ratio corresponds to that of AGN PG0804+761 \citep{Kaspietal2000,2004ApJ...613..682P}. Then we constrain the $R-L$ relations for the two equal-sized subsamples (low- and high-accreting) of 78 measurements (thus excluding one measurement corresponding to PG0804+761) in the fixed flat $\Lambda$CDM model (with $H_0=72\,{\rm km\,s^{-1}\,Mpc^{-1}}$ and $\Omega_{\rm m0}=0.3$).\footnote{When allowing $\Om$ to vary with $H_0=70\,{\rm km\,s^{-1}\,Mpc^{-1}}$, the constraints for the high-accreting subsample are: $\Om > 0.543$ ($1\sigma$), $\gamma = 0.472 \pm 0.043$, $\beta = 1.315 \pm 0.030$, and $\sigma_{\text{int}} = 0.174^{+0.019}_{-0.023}$; consistent with Eq.~\eqref{eq_RL_high}. For the low-accreting subsample: $\Om > 0.220$ ($2\sigma$), $\gamma = 0.469 \pm 0.050$, $\beta = 1.430 \pm 0.049$, and $\sigma_{\text{int}} = 0.221^{+0.022}_{-0.027}$; with $\gamma$ slightly larger than Eq.~\eqref{eq_RL_low}.} The best-fit $R-L$ relations for the complete Best sample as well as the low-accreting and the high-accreting subsamples are shown in Fig.~\ref{fig_RLrel_lowhigh} in the left, middle, and right panels, respectively. The inferred $R-L$ relations are as follows,
\begin{itemize}
    \item the complete Best sample:
    % \begin{align}
    % \log{\left(\frac{\tau}{\text{day}} \right)}&=(0.40^{+0.02}_{-0.02})\log{\left(\frac{L_{5100}}{10^{44}\,{\rm erg\,s^{-1}}}\right)}+(1.37^{+0.02}_{-0.02})\,,\notag\\
    % \sigma_{\rm int}&=0.20^{+0.02}_{-0.01}\,,
    % \label{eq_RL_all}
    % \end{align}
    \begin{equation}
        \label{eq_RL_all}
        \resizebox{0.48\textwidth}{!}{%
        $\begin{aligned}
        \log{\left(\frac{\tau}{\text{day}} \right)} &= (0.40^{+0.02}_{-0.02})\log{\left(\frac{L_{5100}}{10^{44}\,{\rm erg\,s^{-1}}}\right)} + (1.37^{+0.02}_{-0.02})\,, \\
        \sigma_{\rm int} &= 0.20^{+0.02}_{-0.01}\,,
        \end{aligned}$%
        }
    \end{equation}
    \item the low-accreting subsample:
    % \begin{align}
    %  \log{\left(\frac{\tau}{\text{day}} \right)}&=(0.43^{+0.05}_{-0.04})\log{\left(\frac{L_{5100}}{10^{44}\,{\rm erg\,s^{-1}}}\right)}+(1.42^{+0.04}_{-0.04})\,,\notag\\
    % \sigma_{\rm int}&=0.21^{+0.02}_{-0.02}\,,
    %  \label{eq_RL_low}
    % \end{align}
    \begin{equation}
        \label{eq_RL_low}
        \resizebox{0.48\textwidth}{!}{%
        $\begin{aligned}
        \log{\left(\frac{\tau}{\text{day}} \right)} &= (0.43^{+0.05}_{-0.04})\log{\left(\frac{L_{5100}}{10^{44}\,{\rm erg\,s^{-1}}}\right)} + (1.42^{+0.04}_{-0.04})\,, \\ % Removed \notag
        \sigma_{\rm int} &= 0.21^{+0.02}_{-0.02}\,,
        \end{aligned}$%
        }
    \end{equation}
    \item the high-accreting subsample:
    % \begin{align}
    %  \log{\left(\frac{\tau}{\text{day}} \right)}&=(0.48^{+0.04}_{-0.04})\log{\left(\frac{L_{5100}}{10^{44}\,{\rm erg\,s^{-1}}}\right)}+(1.31^{+0.03}_{-0.03})\,,\notag\\
    % \sigma_{\rm int}&=0.17^{+0.02}_{-0.02}\,.
    % \label{eq_RL_high}
    % \end{align}
    \begin{equation}
        \label{eq_RL_high}
        \resizebox{0.48\textwidth}{!}{%
        $\begin{aligned}
        \log{\left(\frac{\tau}{\text{day}} \right)} &= (0.48^{+0.04}_{-0.04})\log{\left(\frac{L_{5100}}{10^{44}\,{\rm erg\,s^{-1}}}\right)} + (1.31^{+0.03}_{-0.03})\,, \\ % Removed \notag
        \sigma_{\rm int} &= 0.17^{+0.02}_{-0.02}\,.
        \end{aligned}$%
        }
    \end{equation}
\end{itemize}
The inferred $R-L$ relations, Eqs.~\eqref{eq_RL_all}, \eqref{eq_RL_low}, and \eqref{eq_RL_high}, show that the intrinsic scatter of the whole sample and both subsamples are comparable, with differences ranging from 0.35$\sigma$ to 1.4$\sigma$. At the same time the smaller subsamples have statistical uncertainties of the inferred $R-L$ relation parameters larger by about a factor of two. Therefore, dividing the sample with respect to the Eddington ratio is not expected to yield significantly tighter cosmological constraints. Generally, it is not appropriate to explicitly use the Eddington ratio as the third parameter in the $R-L$ relation as it artificially enhances the correlation because of the proportionality $\lambda_{\rm Edd}\propto L_{5100}/\tau$ (self-correlation). Instead, independent accretion-rate proxies, such as the flux ratio of the \Feii\ emission line with respect to other broad lines $\mathcal{R}_{\rm FeII}$, should be used. However, the previous inclusion of the flux ratio between the optical \Feii\ and H$\beta$ emission lines $\mathcal{R}_{\rm FeII}$ as the third parameter in the $R-L$ relation did not yield an improvement of $R-L$ relation and cosmological parameter constraints \citep{Khadkaetal2021c}. The same was noted for the flux ratio of the UV \Feii\ and \mii\ emission lines for the case of \mii\ RM QSOs \citep{Khadkaetal2022a}.

In terms of the $R-L$ relation parameters, the low-accreting subsample has a consistent $R-L$ relation with the whole sample within 1$\sigma$ uncertainties. The high-accreting subsample exhibits a steeper slope close to the value of 0.5 based on simple photoionization arguments, while also having a smaller intercept, which is consistent with the previously reported effect of time-delay shortening for high-accreting AGNs due to e.g. a self-shadowing effect of the puffed-up slim disc that obscures the inner part of the photoionizing disc emission \citep{2014ApJ...797...65W}.

In the low-accreting subsample, a shallower $R-L$ slope can be interpreted as being the consequence of a few AGNs that exhibit shorter H$\beta$ lags than is predicted by the photoionization model. This deviation may be explained by retrograde accretion \citep{1972ApJ...178..347B}, where the spin angular momentum of the black hole is opposite to that of the accretion flow \citep[also see discussion in][]{2018ApJ...856....6D}. Such shortened H$\beta$ lags in low-accreting AGNs have also been reported in the SDSS-RM campaigns \citep{2015ApJS..216....4S, 2015ApJ...805...96S, 2017ApJ...851...21G}.

Therefore, our analysis demonstrates that while the H$\beta$ $R-L$ relation remains broadly standardizable across different cosmological models, its parameters depend mildly on the Eddington ratio. The observed deviation from the canonical relation underscores the systematic effect of sample selection. In particular, dividing the AGN sample into distinct accretion rate regimes may help mitigate this effect. Notably, previous studies have found that moderately accreting AGNs tend to follow the canonical relation \citep{2013ApJ...767..149B}, suggesting that the accretion rate plays a key role in the observed scatter.

\section{Conclusion}
\label{sec:conclusion}

In this paper, we test the standardizability of a homogeneous sample of 157 H$\beta$ RM AGN by simultaneously constraining $R-L$ relation and cosmological parameters under six cosmological models/parametrizations. We improve upon previous analyses \citep{Caoetal2025} by including more AGNs and focusing only on H$\beta$ monochromatic measurements. We have found that these H$\beta$ RM AGNs can be standardized using this $R-L$ relation and the posterior cosmological parameter constraints are consistent with, but significantly less restrictive than, those from $H(z)$ + BAO data. Additionally, we find that the scatter in the $R-L$ relation moderately depends on the Eddington ratio, indicating that a more careful selection of the sample is necessary to minimize such systematics when using AGNs as cosmological probes. However, the current sample is still limited in terms of the dynamic range of Eddington ratios, including only a few sources with moderately high and low accretion rates, but no extreme cases. To robustly assess the viability of using H$\beta$ RM AGNs as cosmological probes, a significantly larger and at the same time homogeneous sample is required, specifically one that spans a broader range in both luminosity and Eddington ratios. Future surveys, such as the SDSS-V Black Hole Mapper \citep{SDSS_BH_Mapper_2017arXiv171103234K} and the Vera C. Rubin Observatory Legacy Survey of Space and Time \citep[LSST;][]{Kovacevic:2022msi,2023A&A...675A.163C}, are expected to play a crucial role in building such datasets.

 Moreover, it may be useful to look for better proxies of the BLR photoionizing flux than the standard monochromatic luminosity. AGNs of different relative accretion rates (bolometric luminosities) have different shapes of UV/optical spectral energy distributions, and hence various relative fractions of photoionizing photons. This is not taken into account when adopting optical luminosities at the same frequency at the rest-frame for all the sources, which likely adds to the overall scatter of the $R-L$ relation \citep{2020ApJ...899...73F}. Compensating for the spectral shape differences may help reduce the intrinsic scatter of the $R-L$ relation, which could be beneficial for cosmological applications as well.

\begin{acknowledgments}
A.K.M. acknowledges the support from the European Research Council (ERC) under the European Union’s Horizon 2020 research and innovation program (grant No. 951549). M.Z. acknowledges the OPUS-LAP/GA\v{C}R-LA bilateral project (2021/43/I/ST9/01352/OPUS
22 and GF23-04053L). The computations for this project were partially performed on the Beocat Research Cluster at Kansas State University, which is funded in part by NSF grants CNS-1006860, EPS-1006860, EPS-0919443, ACI-1440548, CHE-1726332, and NIH P20GM113109. %This project has received funding from the European Research Council (ERC) under the European Union’s Horizon 2020 research and innovation program (grant agreement No.~[951549]). BC and MZ ackowledge the OPUS-LAP/GA\v{C}R-LA bilateral project (2021/43/I/ST9/01352/OPUS22 and GF23-04053L).
\end{acknowledgments}

%\appendix

%\section{H$\alpha$ and H$\beta$ RM AGN data}
%\label{HalphaQSO}

% \nocite{*}

\bibliographystyle{apsrev4-2-author-truncate}
\bibliography{apssamp}% Produces the bibliography via BibTeX.

\end{document}